\documentclass[rmp,10pt,twocolumn,showpacs,amsmath,amssymb,floatfix]{revtex4-1}
\usepackage[pdftex]{graphicx} % Include figure files
\usepackage{dcolumn}  % Align table columns on decimal point
\usepackage{bm}       % bold math
\usepackage[usenames,dvipsnames]{color}
\definecolor{URLCOL}{rgb}{0,0.17,0.43} %external link color
\definecolor{LINKCOL}{rgb}{0.05,0.4,0} %internal link color
\definecolor{CITECOL}{rgb}{0.35,0,0.48} %link to bibliography
\definecolor{goodgreen}{RGB}{42,125,35} %better green
\usepackage{epstopdf}
\usepackage[pdftex,bookmarks,breaklinks,bookmarksopen,bookmarksnumbered,colorlinks,linkcolor=LINKCOL,linktocpage,citecolor=CITECOL,urlcolor=URLCOL,pdfpagemode=UseOutline,pdftex]{hyperref}

\usepackage{microtype}

\usepackage{booktabs}
\usepackage{verbatim}
\usepackage{comment}
\usepackage{natbib}
\definecolor{TITLECOL}{rgb}{0.1,0.2,0.7} %title color
\definecolor{PCOL}{rgb}{0.5,0.06,0.01} %title color
\definecolor{CHAPCOL}{rgb}{0,0.48,0} %chapter color
\definecolor{SECOL}{rgb}{0.1,0.2,0.7} %sec color
\definecolor{CONTENTSCOL}{rgb}{0.1,0.2,0.7} %can choose the table of contents title to have same color as sec
\definecolor{SSECOL}{rgb}{0.25,0,0.48} %ssection color
\definecolor{SSSECOL}{rgb}{0.2,0.08,0.53} %subsubsection color  0.2,0.08,0.53
\definecolor{SHDCOL}{rgb}{0.4,0,0} % heading of section color
\definecolor{ITMCOL}{rgb}{0.4,0,0} % bulletted text of item color
\definecolor{EXCOL}{rgb}{0,0.47,0.01} %color of exercises
\definecolor{DEFCOL}{rgb}{0,0.42,0.01} %color in definitions file for definition headings
\def\coloredtitle#1{
\newcommand*{\keeptitle}{#1}
\title{\textcolor{TITLECOL}{#1}}
} %title color

\def\coloredauthor#1{
\newcommand*{\keepauthor}{#1}
\author{\textcolor{CITECOL}{#1}}
} %author color

\def\sec#1{\section{\textcolor{SECOL}{#1}}}
\def\ssec#1{\subsection{\textcolor{SSECOL}{#1}}}

\def\bea{\begin{eqnarray}}
\def\eea{\end{eqnarray}}
\def\ben{\begin{equation}}
\def\een{\end{equation}}
\def\bei{\begin{itemize}}
\def\eei{\end{itemize}}
\def\beit{\begin{itemize}}
\def\eit{\end{itemize}}
\def\benu{\begin{enumerate}}
\def\enu{\end{enumerate}}
\def\n{n}
\def\sss{\scriptscriptstyle\rm}
\def\l{^\lambda}

\def\hatV{{\hat V}}
\def\hatH{{\hat H}}
\def\1var{(\bx_1...\bx\N)}
\def\half{\frac{1}{2}}

\def\br{{\bf r}}

\def\bx{{x}}

\def\cN{{\cal N}}
\def\x{_{\sss X}}
\def\c{_{\sss C}}
\def\s{_{\sss S}}
\def\xc{_{\sss XC}}
\def\Hx{_{\sss HX}}
\def\Hxc{_{\sss HXC}}

\def\xco{_{{\sss XC},1}}
\def\xct{_{{\sss XC},2}}
\def\N{_{\sss N}}
\def\H{_{\sss H}}
\def\W{^{\rm W}}
\def\ext{_{\rm ext}}

\def\HF{^{\rm HF}}
\def\EXX{^{\rm EXX}}

\def\BALDA{^{\rm BALDA}}

\def\HOMO{^{\rm HOMO}}
\def\LUMO{^{\rm LUMO}}

\def\unif{^{\rm unif}}

\def\ee{_{\rm ee}}

\def\up{_\uparrow}
\def\dn{_\downarrow}

\def\sph_int{ {\int d^3 r}}

\def\intr{\int d^3r\,}
\def\intrp{\int d^3r'\,}
\def\dv{\Delta v}
\def\dvs{\Delta v\s}

\begin{document}

\thispagestyle{empty}

\coloredtitle{The Hubbard Dimer: A density functional case study of a many-body problem}

\coloredauthor{D. J. Carrascal$^{1,2}$, J. Ferrer$^{1,2}$, J. C. Smith$^3$ and K. Burke$^3$}

\affiliation{$^1$ Department of Physics, Universidad de Oviedo, 33007 Oviedo, Spain}
\affiliation{$^2$Nanomaterials and Nanotechnology Research Center, Oviedo, Spain}
\affiliation{$^3$ Departments of Chemistry and of Physics, University of California, Irvine, CA 92697,  USA}

\date{\today}

\begin{abstract}

This review explains the relationship between density functional theory and strongly
correlated models using the simplest possible example, the two-site Hubbard 
model.  The relationship to traditional quantum chemistry is included.  
Even in this elementary
example, where the exact ground-state energy and site occupations can be found analytically,
there is much to be explained in terms of the underlying logic and aims of Density Functional Theory.  
Although the usual solution is analytic, the density functional is given only implicitly.
We overcome this difficulty using the Levy-Lieb
construction to create a parametrization of the exact function with negligible errors.
The symmetric case is most commonly studied, but we find a rich variation in behavior
by including asymmetry, as strong correlation physics vies with charge-transfer effects.
We explore the behavior of the gap and the many-body Green's function, demonstrating
the `failure' of the Kohn-Sham method to reproduce the fundamental gap. 
We perform benchmark calculations of the occupation and components
of the KS potentials, the correlation kinetic energies, and the adiabatic connection.
We test several approximate functionals (restricted and unrestricted Hartree-Fock and
Bethe Ansatz Local Density Approximation) to show their successes and limitations.
We also discuss and illustrate the concept of the derivative discontinuity.
Useful appendices include analytic expressions for Density Functional energy components, several limits
of the exact functional (weak- and strong-coupling, symmetric and asymmetric), the  
Kohn-Sham hopping energy functional for 3 sites, various adiabatic connection results,
proofs of exact conditions for this model, and the origin of the Hubbard model from 
a minimal basis model for stretched H$_2$.

\end{abstract}


\maketitle

\def\T{\hat{T}}
\def\V{\hat{V}}
\def\eps{\epsilon}
\def\deps{\Delta\eps}
\def\dn{\Delta\n}
\def\Ts{\sqrt{n_1 n_2}}
\def\Uh{\frac{U}{4}\left(n_1^2 + n_2^2\right)}
\def\bUh{\left(n_1^2 + n_2^2\right)}  %"bare" Hartree, i.e. just the n dependence
\def\dndo{x}
\def\dndd{x^2}
\def\dndf{x^4}
\def\al{\alpha}
\def\gm{\gamma}
 \newcommand{\sgn}{\operatorname{sgn}}
%\tableofcontents

%\sf

\newpage
\sec{Introduction}

In condensed matter, the world of electronic structure theory can be divided into 
two camps:  the weakly and the strongly correlated.   Weakly correlated
solids are almost always treated with density-functional methods
as a starting point for ground-state properties\cite{DG90,Kb99,C06,B12,BW13}.
Many-body (MB) approximations
such as GW might then be applied to find properties of the quasi-particle
spectrum, such as the gap\cite{VGH95,PSG98,AG98}.  
This approach is `first-principles', in the sense that it uses the real-space
Hamiltonian for the electrons in the field of the nuclei, and produces a 
converged result that is independent of the basis set, once a sufficiently
large basis set is used.  Density functional theory (DFT) is known to be exact 
in principle, but the
usual approximations often fail when correlations become strong\cite{CMY08}.

On the other hand, strongly correlated systems are most often treated via
lattice Hamiltonians with relatively few parameters\cite{KE94,Dc94}.
These simplified Hamiltonians
can be easier to deal with, especially when correlations are strong\cite{EKS92,Dc94}.
Even approximate solutions to such Hamiltonians can yield insight into the
physics, especially for extended systems\cite{S08}.  However, such Hamiltonians can
rarely be unambiguously derived from a first-principles starting point, making it difficult
(if not impossible) to say how accurate such solutions are quantitatively or to
improve on that accuracy.  Moreover, methods that yield approximate Green's functions
are often more focused on response properties or thermal properties rather
than on total energies in the ground-state.

On the other hand, the ground-state energy of electrons plays a much more
crucial role in chemical and material science applications\cite{Mc04,PY89}.  Very small energy differences determine
geometries and sometimes qualitative properties, such as the nature of a transition
state in a chemical reaction\cite{LS95,HRJO04,FP07} or where a molecule is adsorbed on a surface\cite{HMN96,OKSW00}.  An error of 0.05 eV changes a reaction rate by a
factor of 5 at room temperature.   Thus quantum chemical development has focused
on extracting extremely accurate energies for the ground and other eigenstates\cite{KBP96,Hb96,F05,S12,YHUM14}.
This is routinely achieved for molecules using coupled-cluster methods (CCSD(T))
and reasonable basis sets\cite{PB82,S97}.  Such methods are called {\em ab initio}, but are not yet
widespread for solids, where quantum Monte Carlo (QMC) is more often 
used\cite{FU96,NU99}.  DFT calculations for molecules
are usually much less computationally demanding, but the errors are less systematic and less reliable\cite{PY95}.  

However, many materials of current technological
interest are both chemically complex
and strongly correlated\cite{B12}.  Numerous metal
oxide materials are relevant
to novel energy technologies, such as
TiO$_2$ for light-harvesting\cite{OG91} or LiO compounds
for batteries\cite{HWC11,TWI12}.  For many cases, DFT calculations find ground-state
structures and parameters, but some 
form of strong correlation method, such 
as introducing a Hubbard $U$ or applying
dynamical mean field theory (DMFT), is needed to correctly
align bands and predict gaps\cite{AAL97,GKKR96}.
There is thus great interest in developing techniques
that use insights from both ends, such as DFT+U and dynamical mean field theory\cite{HFGC14,APKA97,KV04,KSHO06,KM10}.

There are two different approaches to combining DFT with lattice Hamiltonians\cite{CC13}.
In the first, more commonly used, the lattice Hamiltonian is taken as given, and
a density function(al) theory is constructed for that Hamiltonian\cite{GS86}.  We say function(al), not
functional, as the density is now given by a list of occupation numbers, rather
than a continuous function in real space.  The parenthetical reminds us that although everything
is a function, it is analogous to the functionals of real-space DFT.
We will refer to this method as SOFT, i.e.,
site-occupation function(al) theory\cite{SGN95},  although in
the literature it is also known as lattice density 
functional theory\cite{IH10}.  While analogs of the basic theorems of real-space DFT
can be proven such as the Hohenberg-Kohn (HK) theorems and the Levy constrained
search formulation for SOFT, it is by no means clear\cite{H86} how such schemes might
converge to the real-space functionals as more and more orbitals (and hence
parameters) are added.  Alternatively, one may modify efficient solvers of lattice
models so that they can be applied to real-space Hamiltonians (as least in 1-D), 
and use them to explore the nature of the exact functionals and the failures
of present approximations\cite{WSBW12,SWWB12}.
While originally formulated for Hubbard-type lattices, SOFT has been extended and
applied to many different models include quantum-spin chains\cite{AC07}, the Anderson impurity 
model\cite{TP11,CFb12}, the 1-D random Fermi-Hubbard model\cite{XPTT06}, and 
quantum dots\cite{SSDE11}. 

These two approaches are almost orthogonal in philosophy.  In the first, one finds
approximate function(al)s for lattice Hamiltonians, and can then perform Kohn-Sham (KS) DFT
calculations on much larger (and more inhomogeneous) lattice problems\cite{CC05}, but with 
all the usual caveats of DFT treatments (am I looking at interesting physics
or a failure of an uncontrolled approximation?).  
For smaller systems, one can often also compare approximate DFT calculations
with exact results, results which would be prohibitively expensive to calculate
on real-space Hamiltonians.
The dream of lattice models in DFT is that lessons we learn on the lattice can be applied to
real-space calculations and functional developments.  To this end, work has been done on
understanding self-interaction corrections\cite{VC10},  and on wedding 
TDDFT and DMFT methods for application to more complex lattices (e.g. 3-D Hubbard)\cite{KPV11}.
And while it is beyond the scope of this current review, much work has been done on developing
and applying density-matrix functional theory for the 
lattice as well\cite{LPb00,LP02,LP03,LP04,SP11,SP14}.
While such results can be
very interesting, it is often unclear how failures of approximate lattice DFT calculations 
are related to failures of the standard DFT approximations in the real world.

There is much interest in extracting excited-state information from DFT, and time-dependent (TD)
DFT\cite{RG84} has become a very popular first-principles approach\cite{BWG05,U12,MMNG12}.   Because exact
solutions and useful exact conditions are more difficult for TD problems,
there has been considerable research using lattices.
TD-SOFT can be proven for the lattice in much the same way SOFT is proven
from ground-state DFT.  This generalization is worked out carefully in Refs. \cite{T11,FT12}.  
Applications of TD-SOFT typically involve Hubbard chains both with and without various
types of external potentials \cite{AG02,KVOC11,TR14,MRHG14}. However, TD-SOFT has 
also been applied to the dimer to understand the effects of the adiabatic approximation in TD-DFT\cite{FFTA13,FM14,FMb14}, 
strong correlation\cite{TR14}, and TD-LDA results for stretched H$_2$ in real-space\cite{AGR02}.
Unfortunately, we will already fill this article simply discussing the ground-state
SOFT problem, and save the TD case for future work.

\begin{figure}[htb]
\includegraphics[width=1.0\columnwidth]{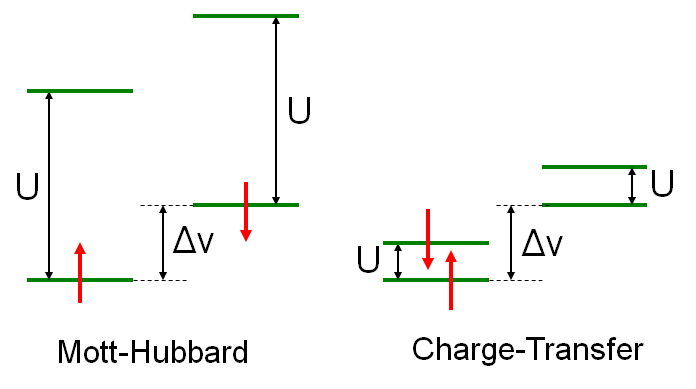}
\caption{Many-body view of two distinct regimes of the asymmetric Hubbard dimer.  On the left,
the charging energy is much greater than the difference in on-site potentials.
On the right, the situation is reversed.}
\label{cart}
\end{figure}
To get the basic idea, consider Fig. \ref{cart}.  It shows the asymmetric Hubbard
dimer in two different regimes.  On the left, the Hubbard $U$ energy is considerably larger
than the difference in on-site potentials and the hopping energy $t$.  This is the case most
often analyzed, where strong correlations drive the system into the Mott-Hubbard regime if $U$ is also 
considerably larger than $t$.  The on-site occupations are in this case close to 1.  
On the right panel, $U$ is in contrast smaller than the on-site potential difference $\dv$, 
and here the dimer stays in the charge-transfer regime, where both electrons mostly sit in the 
same deeper well.
This is the many-body view of the physics of an asymmetric Hubbard dimer.  

\begin{figure}[htb]
\includegraphics[width=0.9\columnwidth]{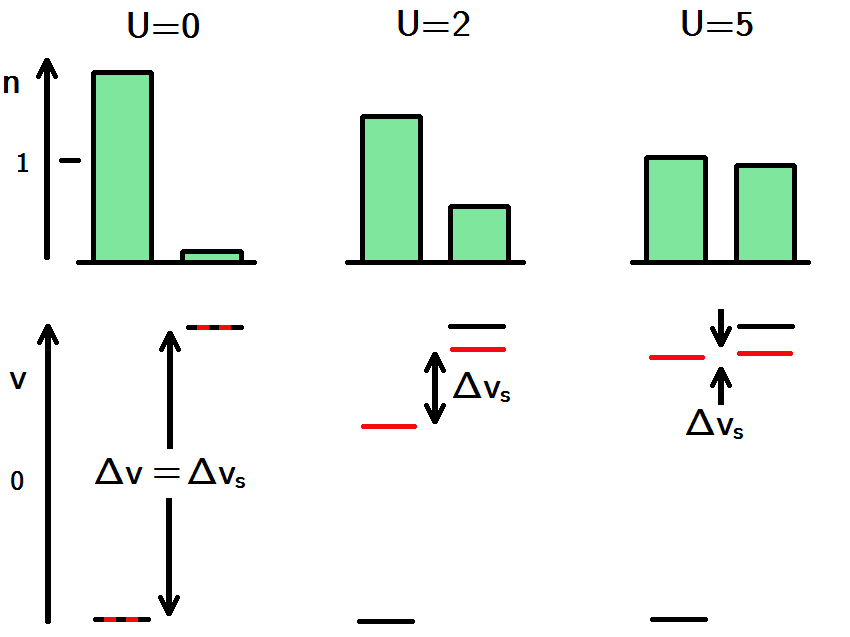}
\caption{Occupations $n$ and potentials $v$ of an asymmetric half-filled Hubbard dimer as a function of $U$.  
The on-site potential difference $\dv$ is shown in black and the KS on-site potential difference $\Delta v\s$ is in red.
The second and third panels correspond to the situations of Fig. \ref{cart}.}
\label{cartB}
\end{figure}
Now we turn to the KS-DFT viewpoint. Here, we replace the interacting Hubbard dimer  ($U\neq 0$)
with a non-interacting ($U=0$) tight-binding dimer, called the KS system, that reproduces the Hubbard occupations.
In Fig. \ref{cartB}, we take the asymmetric dimer with the same on-site potential difference,
but we vary $U$.  We plot the occupations, showing how, as $U$ increases, their difference
decreases.  But we also plot the on-site potentials of the Kohn-Sham model, $\Delta v\s$, that
are chosen to reproduce the occupations of the interacting system with a given value of $U$.
 As $U$ increases, the KS on-site potential difference reduces  and the offset from 0 increases.  The middle
panel corresponds to the charge-transfer conditions of Fig. \ref{cart}, while the last panel
corresponds to the Mott-Hubbard conditions of Fig. \ref{cart}.
The basic theorems
of DFT show that if we know the energy as a function(al) of the density, we can
determine the occupations by solving effective tight-binding equations, the
KS equations, and then find the \emph{exact}
ground-state energy.  
This is not mean-field theory.  It is instead a horribly contorted logical construction,
that is wonderfully practical for computations of ground-state quantities.
Inside this article, we give explicit formulas for the energy functional of
the Hubbard dimer.

We perform a careful study of the
Hubbard dimer, to show the differences between SOFT and real-space DFT.  
We show how  
it is {\em necessary} to introduce inhomogeneity into the site occupations in order to 
find the exact density function(al) explicitly.  
In Section \ref{sec:dft} we explain the logic of the KS DFT approach in excruciating detail in order to both
illustrate the concepts to those unfamiliar with the method and to give explicit
formulas for anyone doing SOFT calculations.
We elucidate the differences between the KS  and the many-body Green's functions in
Section \ref{Green}.
Next, in Sections \ref{gap} and \ref{Correlation} we discuss in detail both concepts and tools for strong correlation, and explain how
the gap problem appears in DFT.
We construct the adiabatic connection
formula for the exact function(al) in Section \ref{sec:AC}, showing how it is quantitatively similar to those of
real-space DFT.  
We use the theory to construct
a simple parameterization for the exact function(al) for this problem in Section \ref{param}, where we also demonstrate
the accuracy of our formula by finding ground-state energies and densities by solving the
KS equations with our parametrization.  
In Section \ref{MF}, we study the broken-symmetry solutions of Hartree-Fock theory,
showing that these correctly yield both the strongly-correlated limit and the approach
to this limit for strong correlation.  
In Section \ref{BALDA} we present BALDA (Bethe-ansatz local density approximation), a popular
approximation for lattice DFT, and in Section \ref{BALDAvsHF} we compare the accuracy of 
BALDA and Hartree-Fock to each other.
We discuss fractional particle number and the derivative discontinuity in Section \ref{frac}.
Finally, we end with a discussion of our results in Section \ref{conclusion}.
In Table \ref{notation} we list our notation for the Hubbard dimer, as well as many standard
DFT definitions.

\begin{table}[htb]
\begin{tabular}{>{$}l<{$}l}
\toprule
\rm{Definition} & \rm{Description}\\ \colrule%\midrule
\rm{Generic~ DFT}\\\colrule
\Psi[n] & Many-body wfn of density $n$\\
\Phi[n] & Kohn-Sham wfn of density $n$\\
F =  T + V\ee  & Hohenberg-Kohn Functional \\
E\xc = F- T\s - U\H  & Exchange-correlation energy \\
E\x =\langle\Phi|\hat{V}\ee|\Phi\rangle - U\H & Exchange energy\\
E\x=-U\H/2  & Exchange energy for 2 electrons \\
E\c = T\c + U\c  & Total correlation energy \\
T\c = T-T\s & Kinetic correlation energy \\
U\c = V\ee - U\H -E\x & Potential correlation energy \\
U\xc(\lambda) = U\xc^{\lambda}/\lambda & Adiabatic connection integrand \\
T\c = E\c - dE\c\l/d\lambda |_{\lambda=1} & Method to extract $T\c$ from $E\c$\\
U\c = dE\c\l/d\lambda|_{\lambda=1} & Method to extract $U\c$ from $E\c$\\
\hat{h}\s = -\nabla^2/2 +v\s & Kohn-Sham hamiltonian\\
v\s = v + v\H +v\xc & Kohn-Sham one-body potential\\
E\c^{\rm{trad}} = E - E^{\rm{HF}} & Quantum chemical corr. energy \\\colrule
\rm{SOFT~Hubbard}\\\colrule
n_1,\,n_2         & Occupations at sites 1, 2 \\
N=n_1+n_2                 & Total number of electrons\\
\Delta n=n_1 -n_2 & Occupation difference \\ 
\Delta m = m_1 - m_2 & Magnetization difference\\
v_1,\,v_2            & On-site potentials \\
\bar{v} = (v_1 + v_2)/2=0 & On-site potential average\\
\Delta v = v_2 - v_1 & On-site potential difference\\ 
\Delta v\xc = v\xct - v\xco & XC potential difference\\
U\H = U(N^2+\Delta n^2)/4 & Hartree energy\\
E\Hx = U(N^2+\Delta n^2)/8 & Hartree-Exchange energy\\
T\s=\!-t\sqrt{(2\!-\!|N-2|)^2-\!\Delta n^2} & Single particle hopping energy \\\colrule
\rm{Dimensionless~Variables}\\\colrule
\epsilon = E/2\,t & Energy in units of hopping\\
u=U/2\,t & Hubbard $U$ in units of hopping\\
\nu = \Delta v /2\,t & Pot. diff. in units of hopping\\
\rho = \Delta n/2 & Reduced density difference\\
\bar{\rho} = 1-\rho & Asymmetry parameter\\
\botrule%\bottomrule
\end{tabular}
\caption{Standard DFT definitions and our Hubbard dimer notation.}
\label{notation}
\end{table}

Our purpose here is several-fold.   Perhaps most importantly, this article is intended to
explain the logic of modern DFT to our friends who are more 
familiar with strongly correlated systems.   We take the simplest model of
strong correlation, and illustrate many of the basic techniques of modern DFT.   
There are many more tricks and constructions, but we save those for future work.
The article should be equally useful to researchers in other fields who are
unfamiliar with the logic of DFT, such as traditional quantum chemists or
atomic and molecular physicists.

Secondly, the article forms an essential reference for those researchers interested
in SOFT, possibly in very different contexts and applied to very different models.
It shows precisely how concepts from first-principles calculations
are realized in lattice models.  Third, we give many exact results for
this simple model, expanding in many different limits, showing that even in this simple
case, there are orders-of-limits issues.  Fourth, we use DFT techniques to find a
simple but extremely accurate parametrization of the exact function(al) for this 
model.  Even though the model can be solved analytically, the function(al) cannot
be expressed explicitly.   Thus our parametrization provides an ultra-convenient
and ultra-accurate expression for the exact function(al) for this model, that can
be used in the ever increasing applications of SOFT.  Finally, we examine several
standard approximations to SOFT, including both restricted and unrestricted 
mean field theory, and the BALDA, and we find
surprising results.

\sec{Background}

In this section we briefly introduce real-space DFT, and the logical underpinnings for everything that follows.
Then we discuss the mean-field approach to the Hubbard model as well as a few well-known results and limits
for the Hubbard dimer. Throughout this section we use atomic units for all real-space
expressions so all energies are in Hartree and all distances are in Bohr.

\ssec{Density functional theory}
\label{sec:dft}
We restrict ourselves to non-relativistic systems within the Born-Oppenheimer approximation
with collinear magnetic fields\cite{ED11}.
Density functional theory is concerned with efficient methods for finding
the ground-state energy and density of $N$ electrons whose Hamiltonian contains
three contributions:
\ben
\label{hamiltonian}
\hat H = \hat T + \hat V\ee + \hat V.
\een
The first of these is the kinetic energy operator, the second is the 
electron-electron repulsion, while the last is the one-body potential,
\ben
\hat V = \sum_{i=1}^N v(\br_i).
\een
Only $N$ and $v(\br)$ change from one system to another, be they atoms,
molecules or solids.  In 1964, Hohenberg and Kohn proved that for a
given electron-electron interaction, there was
at most one $v(\br)$ that could give rise to the ground-state one-particle 
density $n_0(\br)$ of the system, thereby showing that all ground-state properties
of that system were uniquely determined by $n_0(\br)$~\cite{HK64}.  The ground-state energy $E_0$
could then be found by splitting the variational principle into two steps via the Levy-Lieb 
constrained search approach\cite{L79,L83}. First, the universal functional $F$
is determined, 
\ben
\label{levy1}
F[\n] = \min_{\Psi\to\n} \langle \Psi |\, \hat T + \hat V\ee\, |\Psi \rangle = T[n]+V\ee[n]
\een
where the minimization is over all normalized, antisymmetric $\Psi$ with
one-particle density $\n(\br)$. This establishes a one-to-one connection between wavefunctions and ground-state
densities,
and enables us to define the minimizing wavefunction
functional $\Psi[n_0]$. Then the ground-state energy is determined by a second 
minimization step of the energy functional $E[n]$,
\ben
\label{levy2}
E_0 =\min_n\left\{E[n]\right\}= \min_\n \left\{ F[\n] + \int d^3r\, \n(\br) \,v(\br)\right\}.
\een
This shows that $E_0$ can be found from a 
search over one-particle densities $n(\br)$ instead of many-body wavefunctions $\Psi$, provided that the functional $F[n]$ is known.
The Euler equation corresponding to the above minimization for fixed $N$ is
simply
\ben
\label{consearch}
\left.\frac{\delta F[\n]}{\delta\n(\br)} \right|_{\n_0(\br)} = - v(\br).
\een
Armed with the exact $F[\n]$, the solution of this equation yields the exact ground-state density which,
when inserted back into $F[\n]$, yields the exact ground-state energy.

To increase accuracy and construct $F[\n]$, 
modern DFT calculations use the Kohn-Sham (KS) scheme that imagines a fictitious
set of non-interacting electrons with the same ground-state density as the real Hamiltonian\cite{KS65}.
These electrons satisfy the KS equations:
\ben
\label{KSeq}
\left\{ -\half \nabla^2 + v\s(\br) \right\}\, \phi_i(\br) = \epsilon_i\, \phi_i(\br),
\een
where $v\s(\br)$ is defined as the unique potential that generates single-electron orbitals
$\phi_i(\br)$ that reproduce the ground-state density of the real system,
\ben
\n_0(\br)=\sum_{occ}\,|\phi_i(r)|^2.
\een
To relate these to the interacting system, we write
\ben
\label{ksF}
F[\n]=T\s[\n]+U\H[\n]+E\xc[\n].
\een
$T\s$ is the non-interacting (or KS) kinetic energy, given by
\ben
T_s[\n] = \half \int d^3\, \sum_{i=1}^N | \nabla \phi_i(\br)|^2
=  \min_{\Phi\to\n} \langle \Phi |\, \hat T  |\Phi \rangle,
\een
where we have assumed the KS wavefunction (as is almost always the case) is a
single Slater
determinant $\Phi$ of single-electron orbitals. 
The second expression follows from Eq. (\ref{levy1}) applied to the KS system,
it emphasizes that $T\s$ is a functional of $\n(\br)$, and the minimizer
defines $\Phi[\n_0]$, the KS wavefunction as a density functional.
Then $U\H[\n]$ is the classical electrostatic self-repulsion of $n(\br)$, 
\ben
U\H[n] = \frac{1}{2} \intr \intrp \frac{n(\br)\, n(\br')}{|\br -\br'|},
\label{Hartree}
\een
and $E\xc$ is called the exchange-correlation
energy, and is {\em defined} by Eq. (\ref{ksF}).   

Lastly, we differentiate Eq. (\ref{ksF}) with respect to the density.  Applying Eq. (\ref{consearch})
to the KS system tells us
\ben
v\s(\br)=-\frac{\delta T\s[\n]}{\delta\n(\br)},
\label{vsdef}
\een
yielding
\ben
v\s(\br)=v(\br) + v\H(\br) + v\xc(\br)
\een
where $v\H(\br)$ is the classical electrostatic potential and 
\ben
v\xc(\br)=\frac{\delta E\xc}{\delta\n(\br)}
\een
is the exchange-correlation potential.
This is the single most important result in DFT, as it closes the set of
KS equations.  Given any expression for $E\xc$ in terms of $\n_0(\br)$,
either approximate or exact, the KS equations can be solved self-consistently
to find $\n_0(\br)$ for a given $v(\br)$.

However, we also note that, just as in all such schemes, the energy of the
KS electrons {\em does} not match that of the real system.   This `KS energy'
is 
\ben
\label{ksorbsum}
E\s[n]=\sum_i \epsilon_i = T\s + V\s,
\een
but the actual energy is
\ben
\label{energydiff}
E_0 = F[\n_0] + V[\n_0] = T\s[\n_0]+U\H[\n_0]+E\xc[\n_0]+V[\n_0]
\een
where $\n_0(\br)$ and $T\s[\n_0]$ have been found by solving the KS equations,
and inserted into this expression.  Thus, in terms of the KS orbital energies,
there are double-counting corrections, which can be deduced from Eqs. (\ref{ksorbsum}) and (\ref{energydiff}):
\ben
\label{orbsum}
E_0= E_s - U\H[\n_0] + E\xc[\n_0] - \intr \n_0(\br)\, v\xc[\n_0](\br).
\een
We emphasize that, with the {\em exact} $E\xc[\n_0]$, solution of the KS equations
yields the {\em exact} ground-state density and energy, and this has been done
explicitly in model cases\cite{WSBW13}, but is computationally exorbitant.  The
practical use of the KS scheme is that simple, physically motivated approximations
to $E\xc[\n_0]$ often yield usefully accurate results for $E_0$, bypassing direct
solution of the many-electron problem.

For the remainder of this article, we drop the subscript 0 for notational
convenience, and energies will be assumed to be ground-state energies, unless
otherwise noted.
For many purposes, it is convenient to split $E\xc$ into a sum of exchange and
correlation contributions.
The definition of the KS exchange energy is simply
\ben
E\x[\n] = \langle \Phi[\n] | \hat V\ee | \Phi[\n] \rangle - U\H[\n],
\een
The remainder is the correlation energy functional
\ben
\label{dftcorr}
E\c[\n]=F[\n] -  \langle \Phi[\n] |\, \hat T + \hat V\ee\, | \Phi[\n] \rangle,
\een
which can be decomposed into kinetic $T\c$  and potential $U\c$ contributions
(see Eqs. (\ref{kincorr}) and (\ref{potcorr}) in Sec. \ref{Correlation}).
Additionally, all practical calculations generalize the 
preceding formulas for arbitrary spin using spin-DFT \cite{BH72}.

For just one particle ($N=1$), there is no electron-electron
repulsion, i.e., $V\ee=0$.  This means
\ben
E\x=-U\H,~~~~E\c=0,~~~~~~(N=1),
\een
i.e., the self-exchange energy exactly cancels the Hartree self-repulsion.
Since there is no interaction, $F^0[\n]=T[\n]=T\s[\n]$, and for one electron
we know the explicit functional:
\ben
\label{vonWeisacker}
T\s = T\W = \int d^3r\, |\nabla\n|^2/(8\n),
\een
which is called the von Weisacker functional\cite{W35}.  
For two electrons in a singlet ($N=2$), 
\ben
E\x=-U\H/2,~~~~~~T\s=T\W,~~~~~~~~~(N=2),
\een
but the correlation components are non-zero and non-trivial.

Many popular forms of approximation exist for $E\xc[\n]$, the most common being
the local density approximation (LDA)\cite{KS65,BH72,PW92}, the generalized gradient approximation
(GGA)\cite{P86,B88,LYP88,PCVJ92,PBE96}, and hybrids of GGA with exact exchange from a Hartree-Fock calculation\cite{B93,PEB96,AB99,HSE06}.
The computational ease of DFT calculations relative to more accurate wavefunction
methods usually allows much larger systems to be calculated, leading to DFT's immense popularity
today\cite{PGB15}.  However, all these approximations fail in the paradigm case of
stretched $H_2$, the simplest example of a strongly correlated system\cite{B01,HCRR14}.

\ssec{The Hubbard model}
\label{sec:hubbard}

\def\K{\hat{T}}
\def\O{\hat{V}}
\def\eps{\epsilon}
\def\deps{\Delta\eps}
\def\dn{\Delta\n}
\def\al{\alpha}
\def\gm{\gamma}

The Hubbard Hamiltonian is possibly the most studied, and simplest, model of a strongly correlated electron system. 
It was initially introduced to describe the electronic properties of narrow-band metals, whose conduction bands are formed 
by $d$ and $f$ orbitals, so that electronic correlations become important\cite{H63,F13}. 
The model was used to describe ferromagnetic, antiferromagnetic
and spin-spiral instabilities and phases, as well as the metal-insulator transition in metals and oxides, 
including high-T$_c$ superconductors\cite{Dc94,LNW06}. 
The Hubbard model is both a qualitative version of a physical system depending on what terms
are built in\cite{A87,Sb90} and also a testing-ground for new techniques since the simpler forms of the Hubbard
model are understood very well\cite{Hb89,BSb89,BSW89,H93}.

The model assumes that each atom in the lattice has a single orbital.
The Hamiltonian is typically written as \cite{M93,Hc96,EFGK05,Te05}
\ben
\hat{H}=\sum_{i,\sigma}\,v_{i\sigma}\,\hat{n}_{i\sigma}-\sum_{i\,j\,\sigma}\,\left(t_{ij}\,{\hat{c}}_{i\,\sigma}^\dagger\,{\hat{c}}_{j\,\sigma}+h.c.\right) +\sum_{i}\,U_i\,{\hat{n}}_{i\uparrow}\,{\hat{n}}_{i\downarrow}
\een
where at its simplest the on-site energies are all equal $v_{i\sigma}=0$ as well as the 
Coulomb integrals $U_i=U$.  Further, the hopping integrals $t_{ij}$ typically couple only
nearest neighbor atoms and are equal to a single value $t$.  

We note that here the interaction is of ultra-short range, so that two electrons only interact if they are on the
same lattice site. Further, they must have opposite spins to obey the Pauli principle.
Simple examples of building in more complicated physics include using next-nearest-neighbor 
hoppings or nearest neighbors Coulomb integrals for high-T$_c$ cuprate calculations and 
magnetic properties\cite{LH87,DM95,DN98}, and varying  on-site potentials used to model 
confining potentials\cite{RNKH08}. Also, adding more orbitals per site delivers multi-band 
Hubbard models, where Coulomb correlations may be added to some or all of the orbitals.
The Hubbard model has an analytical solution in one dimension, via Bethe ansatz techniques\cite{LW68,LW03}.

If the Hubbard $U$ is small enough, a paramagnetic mean-field (MF)
solution provides a reasonable description of the model in dimensions equal 
or higher than two. As an example, the Hubbard model in a honeycomb
lattice can describe correctly a number of features of gated graphene samples\cite{H06}. 
However, for large $U$ or in one dimension, more sophisticated approaches
are demanded, which go beyond the scope of this article\cite{LW68,F13}. 

We describe briefly the well-known broken-symmetry MF solution,
where the populations of up- and down-spin electrons can differ. 
The standard starting point for the MF solution 
neglects completely quantum fluctuations:
\ben
\label{MFdef}
\left({\hat{n}}_{i\uparrow}-n_{i\uparrow}\right)\,\left({\hat{n}}_{i\downarrow}-n_{i\downarrow}\right)=0, ~~~~~~~~~ (MF)
\een
where $n_{i\sigma} = \langle \hat{n}_{i\sigma} \rangle$, so that 
\ben
{\hat{V}}_{ee}^{MF}=\sum_{i}\,U\,\left(n_{i\uparrow}\,{\hat{n}}_{i\downarrow}+n_{i\downarrow}\,\hat{n}_{i\uparrow}-n_{i\uparrow}\,n_{i\downarrow}\right).
\een
The MF hamiltonian is then just an effective single-particle problem
\bea
\hat{H}^{MF}&=&\sum_{i\sigma} \hat{h}^{\rm{eff}}_{i\sigma},\\
\label{MF_hamiltonian}
\hat{h}^{\rm{eff}}_{i\sigma} &=& v_{i\sigma}^{MF}\,\hat{n}_{i\sigma}-t\,\sum_{j}(\hat{c}_{i\sigma}^{\dagger}\hat{c}_{j\sigma}+h.c.),
\eea
where $v_{i\sigma}^{MF}=v_{i\sigma}+U\, n_{i\bar{\sigma}}$. This $\hat{H}^{MF}$
can be easily diagonalized if one assumes space-homogeneity of
the occupations $n_{i,\sigma}=n_\sigma$.
For large $U$, the broken symmetry solution (often ferromagnetic) has lower energy than the
paramagnetic solution.

\ssec{The two-site Hubbard model}
\def\vo{v_1}
\def\vt{v_2}
\def\no{n_1}
\def\nt{n_2}
\def\deps{\Delta v}
\def\dn{\Delta \n}
\def\dm{\Delta m}
\def\r{W}
\def\y{y}
We now specialize to a simple Hubbard dimer model with open boundaries, but
we allow different on-site spin-independent energies by introducing a third term that produces
asymmetric occupations,
\begin{equation}
\hatH = -t\, \sum_{\sigma} \: (\hat{c}_{1\sigma}^{\dagger}\hat{c}_{2\sigma} + h.c) +
U \sum_{i} \hat{n}_{i\uparrow\,}\hat{n}_{i\downarrow} + \sum_i v_i \hat{n}_i
\end{equation}
where we have made the choices $t_{12}=t^*_{21}=t$ and $\vo + \vt =0$. Our notation for this
Hamiltonian can be found in Table \ref{notation}.

\begin{figure}[htb] 
\includegraphics[width=0.8\columnwidth]{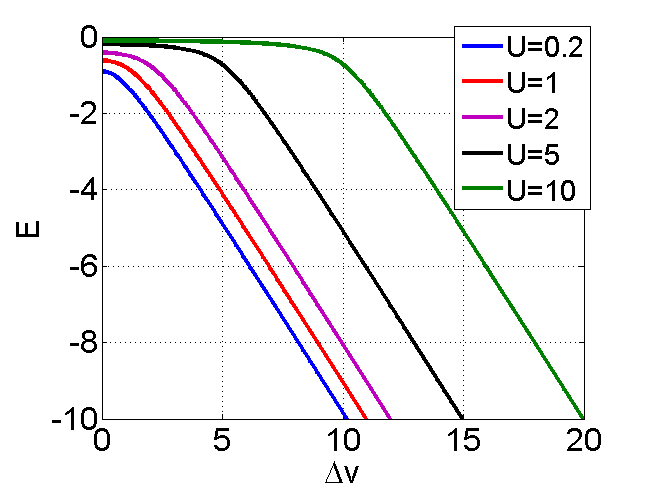}
\caption{Ground-state energy of Hubbard dimer as a function of $\deps$ for several values 
of $U$ and $2\,t=1$.}
\label{gsenergy}
\end{figure}
\begin{figure}[htb] 
\includegraphics[width=0.8\columnwidth]{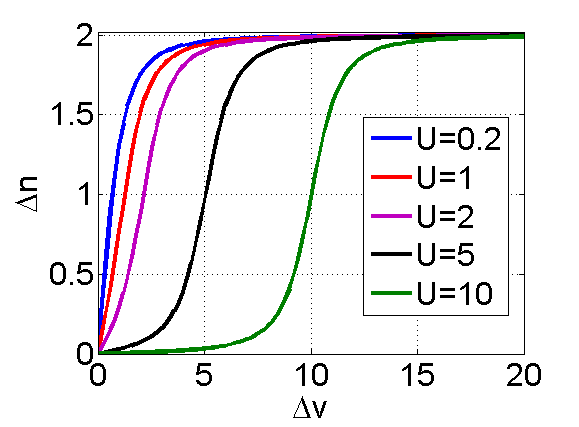}
\caption{Ground-state occupation of Hubbard dimer as a function of $\deps$ for several values 
of $U$ and $2\,t=1$.}
\label{dndv}
\end{figure}

 
It is straightforward to find an analytic solution of the model for any integer occupation $N$. 
However, we specialize to the particle sub-space $N=2$, $S_z=0$ in what follows unless otherwise stated.
We expand the Hamiltonian in the basis set $\left[
|1\uparrow\,1\downarrow\},|1\uparrow\,2\downarrow\},|1\downarrow\,2\uparrow\},
|2\uparrow\,2\downarrow\}\right]$:
\ben
\hat{H}=\left(\begin{array}{cccc}
2\vo+U & -t & t & 0\\
-t & 0 & 0 & -t\\
t & 0 & 0 & t\\
0 & -t & t & 2\vt+U
\end{array}\right)
\een
The eigenstates are three singlets and a
triplet state. The ground-state energy corresponds to the lowest-energy singlet, and 
can be found analytically.  The expressions are given in appendix \ref{energycomp}.
The wavefunction, density difference, and individual energy components are also
given there.
We plot in Fig. \ref{gsenergy} the ground-state energy as a function of $\deps$ for 
several values  of $U$, while in Fig. \ref{dndv}, we plot the occupations.  

When $U=0$, we have the simple tight-binding result, for which
the ground-state energy is
\bea
\label{e0_N2_U0}
E&=&-\sqrt{(2\,t)^2+\deps^2} ~~~~~~~~~~~~~~~~~ (U=0),\\
\label{dn_N2_U0}
\dn&=&2\,\deps/\sqrt{(2\,t)^2+\deps^2}~~~~~~~~~~~ (U=0).
\eea
where $\dn$ is defined in Table \ref{notation}. If there is only one electron, these become smaller by a factor of 2.
The curves for $U=0.2$ are indistinguishable (by eye) from the tight-binding result.
We may simplify the expressions by introducing an effective hopping parameter,
\def\tt{\tilde t}
\ben
\tt = t \sqrt{1+ ({\dv}/({2\,t}))^2}
\label{tildet}
\een
which accounts for the asymmetric potential.  Then 
\bea
E&=&-2\tt ~~~~~~~~~~~~~~~~~ (U=0),\\
\dn&=&\deps/\tt~~~~~~~~~~~~~~~ (U=0),
\eea
i.e., the same equations as when $\dv=0$.

In the other extreme, as $U$ grows,
we approach the strongly correlated limit.
For a given $\dv$, as $U$ increases, $\dn$ decreases as in 
Figs. \ref{cartB} and \ref{dndv}, and
the magnitude of the energy shrinks.
Typically, the $E(\deps)$ curve morphs from
the tight-binding result towards two straight lines for $U$ large:
\bea
E \simeq (U-\deps)\,\Theta(\deps-U) ~~~~~~~~~~~~~~~ U \gg 2\,t\\
\dn \simeq 2\,\Theta(\deps-U) ~~~~~~~~~~~~~~~ U \gg 2\,t
\eea
We also have a simple well-known result for the symmetric limit, $\Delta v$=0, where
\ben
E=-\sqrt{(2 t)^2+(U/2)^2}+U/2 ~~~~~~~~~~ (\dn=\deps =0)
\een
This vanishes rapidly with $1/U$ for large $U$.  Its behavior is different from 
the case with finite $\dv$.
Results for various limits and energy components are given in Appendix \ref{energycomp}.

\ssec{Quantum chemistry}
\label{quanchem}

\def\xv{x} %\deps/2\,t
\def\xmf{x_{MF}} %\def\xmf{x_{\rm{MF}}}
\def\xs{x\s}
\def\eff{^{\rm eff}}

\begin{figure}[htb]
\includegraphics[width=0.8\columnwidth]{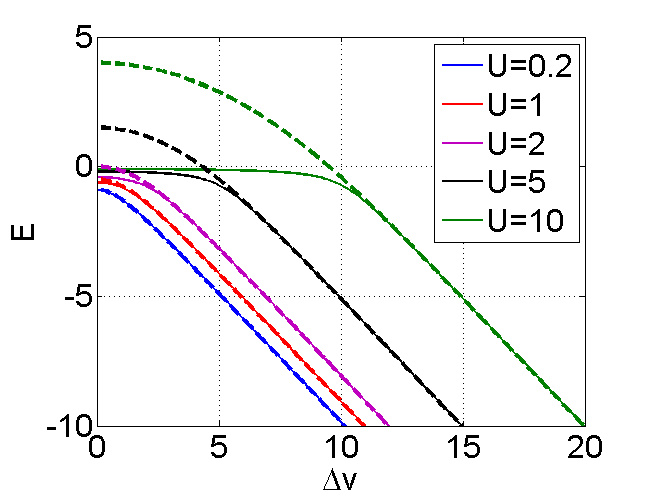}
\caption{Ground-state energy of the Hartree-Fock Hubbard dimer (thick dashed line) and exact 
ground-state of the Hubbard dimer (thin solid line) as a function of $\dv$ for
several values of $U$ and $2\,t=1$.}
\label{HFgs}
\end{figure}

Traditional quantum chemical methods (often referred to as {\em ab initio} by
their adherents) usually begin with the solution of the Hartree-Fock equations\cite{SO82}.
For our Hubbard dimer, these are nothing but the mean-field equations of Sec \ref{sec:hubbard}.
Expressing the paramagnetic HF Hamiltonian of Eq. (\ref{MF_hamiltonian}) for two sites yields a 
simple tight-binding Hamiltonian and eigenvalue equation describing a single-particle in an 
effective potential:
\ben
\label{MFeff}
v\eff_i(n_i) = v_i + U n_i /2.
\een
with an eigenvalue:
\ben
\epsilon\eff = \left(U - \sqrt{(\deps\eff)^2 + (2\,t)^2}\right)/2.
\label{MForb}
\een
Writing
$\phi\eff= (c_1,c_2)^T$, then  
\ben
\label{MFdn}
\dn = 2\, (c_1^2 - c_2^2)= 2\, \frac{1- \xi^2}{1+\xi^2},
\een
where $x=\dv\eff/2\,t$, and
$\xi = \sqrt{x^2 +1} -x$.
Eq. (\ref{MFdn}) is quartic in $\dn$ and can be solved algebraically to find $\dn$ as a 
function of $\deps$ explicitly (appendix \ref{app:MF}).
Just as in KS, the HF energy is not simply twice the orbital energy, there is
a double-counting correction:
\bea
E^{MF}&=& 2\epsilon\eff - U\H \\
&=& \frac{U}{2}\left(1-\left(\frac{\dn}{2}\right)^2\right) -2\,t \sqrt{1 + x^2}.
\nonumber
\eea
These energies are plotted in Fig \ref{HFgs}.
We see that for small $U$, HF is very accurate, but much less so for $2\,t \ll U \ll \dv$.
In fact, the HF energy becomes positive in this region, unlike the
exact energy, which we prove is never positive in appendix \ref{proofs}.  The molecular orbitals often used in chemical descriptions have
traditionally been those of HF calculations, despite the fact that HF energies
are usually far too inaccurate for most chemical energetics\cite{BB00}.
(They have now largely been supplanted by KS orbitals.)
In quantum chemical language, the paramagnetic mean-field solution is called restricted HF (RHF) because
the spin symmetry is restricted to that of the exact solution, i.e., $S_z=0$.  For large enough
$U$, the broken-symmetry, or unrestricted, solution is lower, and is labeled UHF, which
we discuss in Sec. \ref{MF}.

\def\trad{^{\rm trad}}
\begin{figure}[htb] 
\includegraphics[width=0.8\columnwidth]{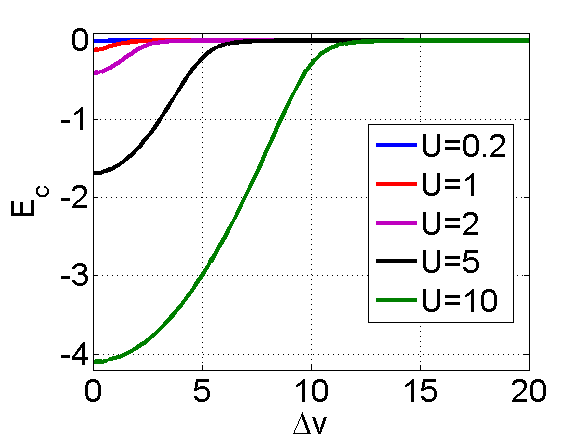}
\caption{Correlation energy $E\c\trad$ of Hubbard dimer as a function of $\deps$ for 
several values of $U$ and $2\,t=1$.}
\label{correnergy}
\end{figure} 
Accurate ground-state energies, especially as a function of nuclear
positions, are central quantities in chemical electronic structure
calculations\cite{SO82}.  Most such systems are weakly correlated
unless the bonds are stretched.
The correlation energy of traditional quantum chemistry is defined as just the error
made by the (restricted) HF solution:
\ben
E\c\trad = E - E\HF.
\label{Ectrad}
\een
This is plotted in Fig \ref{correnergy}.
This is always negative, by the variational principle.  Many techniques have been highly
developed over the decades to go beyond HF.  These are
called model chemistries, and
for many small molecules, errors in energy differences of
less than 1 kcal/mol (0.05 eV) are now routine\cite{OPW95,BM07}.

Usually $E\c\trad$ is a small fraction of $E$ for weakly correlated systems. For example, for the
He atom, $E = -77.5$ eV, but $E\c\trad = -1.143$ eV.  This is the error made by a HF calculation.  In
Fig. \ref{correnergy} we plot $E\c\trad$ just as we plotted $E$ in Fig. \ref{HFgs}. We see that for strong correlation
$E\c\trad$ becomes large ($\sim - U/2$ for $\dv \ll U$), much larger than $E$.
However, $E$ is much smaller, and so any strongly correlated method should reproduce
$E$ accurately.  In fact, one can already see difficulties for weakly correlated
approximations in this limit.  For weak correlation, a small percent error in $E\c\trad$ yields
a very small error in $E$, but produces an enormous error in $E$ in the strong correlation limit.
For an infinitely stretched molecular bond, $t\to 0$ while $U$ remains
finite, so only one
electron is on each site.  Thus $E \to 0$, so we can think
of $E$ as the ground-state electronic energy relative to the dissociated limit, i.e. the binding
energy.

Because HF is accurate when correlation is weak, and because quantum chemistry focuses 
on energy differences, the error is often measured in terms of the accuracy of the
exchange-correlation together (if both are approximated as in most DFT calculations).  
For 2 electrons having $S_z=0$, the
exact exchange is trivial, and so we will focus on approximations to the correlation energy.

Notice the slight difference in definition of correlation energy between DFT  (Eq. \ref{dftcorr}) and quantum
chemistry (Eq. (\ref{Ectrad}))\cite{SL86,GPG96,UG94}.  In DFT, all quantities are defined on a given density, usually
the exact density of the problem, whereas in quantum chemistry, the HF energy is evaluated on the density
that minimizes the HF energy.  For weakly correlated systems, this difference is extremely 
small\cite{GE95}, but is not so small for large $U$. 
And, one can prove, $E\c^{\rm{trad}} \geq E\c^{\rm{DFT}}$\cite{GPG96}, (see appendix \ref{proofs}).

We close by emphasizing the crucial difference in philosophy between DFT and traditional approaches.  In many-body
theory, mean-field theory is an approximation to the many-body problem, yielding an approximate
wavefunction and energy which are expected to be reasonably accurate for small $U$.  In DFT,
this treatment arises from approximating $F$ for small $U$, and so should yield an accurate
KS wavefunction and expectation values for small $U$.  Thus, only one-body properties that
depend only on position are expected to be accurate, and their accuracy can be improved by
further improving the approximation to $F$. For large $U$, such an approximation fails, but there
is still an exact $F$ that yields an exact answer.

\sec{Site-occupation function theory (SOFT)}

In this section, we introduce the site-occupation function
theory for the Hubbard dimer\cite{GS86,SG88,SGN95,CFb12}.   
If we want a physical system where this arises, think of stretched H$_2$\cite{MWY60}.
We imagine a minimal basis set of one function
per atom for the real Hamiltonian.  We choose these basis functions to
be $1s$ orbitals centered on each nucleus, but symmetrically orthonormalized.
Then each operator in real-space contributes to the parameters in the
Hubbard Hamiltonian as seen in Appendix \ref{app:cnxn}.

It is reasonably straightforward to establish the validity of SOFT for
our dimer.  So long as each occupation can come from only one value of $\Delta v$,
for a fixed $U$, there is a one-to-one correspondence between
$\dn$ and $\dv$, and all the usual logic of DFT follows.  But
note that $\K$ and $\O$ in SOFT do {\em not} correspond to the real-space
kinetic energy and potential energy.   For example, the hopping energy
is negative, whereas the real-space kinetic energy is positive. 
This means that all theorems of DFT to be used must be reproven for the
lattice model.  More importantly, the SOFT does not become
real-space DFT in some limit of complete basis sets 
(in any obvious way).  We will however apply the same logic as real-space DFT,
with the hopping energy in SOFT playing the role of the kinetic energy in 
DFT, and the on-site energy in SOFT playing the role of the one-body potential.
The interaction term obviously plays the role of $\hat V\ee$.

\ssec{Non-interacting warm-up exercise}

To show how SOFT works, begin with the $U=0$ case, i.e., tight-binding of
two non-interacting electrons.  The ground-state is always a spin
singlet.  From the non-interacting solution, we can 
solve for $\deps$ in terms of $\dn $
\ben
\label{TBdv}
\deps = \frac{2\,t\,\dn}{\sqrt{4 - \dn^2}},
\een
and substitute back into the kinetic energy expectation value to find
\ben
\label{tsdimer}
T(\no,\nt)=-2\,t\,\Ts. 
\een
This is the universal density function(al) for this non-interacting problem (see Eq. (\ref{levy1})),
and can be used to solve every non-interacting dimer.

To solve this $N=2$ problem in the DFT way, we note that $T$ is playing the role of $F(\no,\nt)$.
So the exact function(al) here is
\begin{equation}
\label{f_n2}
F(\no) = -2\,t\Ts, ~~~~~~~~~~~~~(U=0)
\end{equation}
from which we can calculate all the quantities of interest using a DFT treatment.  
Note that everything is simply a function(al) of $\no$ since $\nt = (N -\no)$, or
alternatively a function(al) of $\dn$.

We then construct the total energy function(al):
\ben 
E(\no)=F(\no)-\deps\,\dn/2, ~~~~~~~~~~~~~~ (U=0)
\een 
and minimize 
with respect to $\no$ for a given $\deps$ to find the ground-state energy
and density:
\bea
\label{e0_n2}
E&=&-\sqrt{(2\,t)^2+\deps^2},\\
\label{n_n2}
\dn&=&2\,\deps/\sqrt{(2\,t)^2+\deps^2}.
\eea
Both of these agree with the traditional approach
and recover Eqs. (\ref{e0_N2_U0}) and (\ref{dn_N2_U0}).  The $N=1$ result 
is half as great as Eqs. (\ref{e0_n2}) and (\ref{n_n2}).

We can deduce several important lessons from this example.
First, we need to vary the one-body potential (in this case, the
on-site energy difference) to make the density change through all possible
values, in order to find the function(al), since it requires knowing
the one-to-one correspondence for all possible densities.  Second, 
if we really change the atoms in our 2-electron stretched molecule,
of course the minimal basis functions would change, and {\em both}
$t$ and $\deps$ would differ.  But here we keep $t$ fixed, and vary 
$\deps$ simply to explore the function(al), even if we are only interested
in solving the symmetric problem.  
(Real-space DFT does not suffer from this problem, as the kinetic and repulsion operators are universal.)
Third, we are reminded that 
the hopping and on-site operators in no sense represent the actual kinetic
and one-body potential terms -- they are a mixture of each.
Finally, although we `cheated' and {\em extracted} the kinetic
energy function(al) from knowing the solutions, if someone had given us the formula,
it would allow us to solve every possible non-interacting Hubbard dimer by minimizing
over densities. And an approximation to that formula would yield approximate solutions
to all those problems.

\ssec{The interacting functional}

For the interacting case, we cannot analytically write down the exact function(al) $F(\no)$
at $N=2$ in closed form.
Although we have analytic formulas for both $E$ and $\dn$ as functions of $\deps$,
the latter cannot be explicitly inverted to yield an analytic formula for $F(\dn)$.
However, we can plot the function(al), by simply plotting  $F=E-V$ as a function of $\no$,
and see how it evolves from  the $U=0$ case to stronger interaction.
The spin state is always a singlet.
\begin{figure}[htb] 
\includegraphics[width=.8\columnwidth]{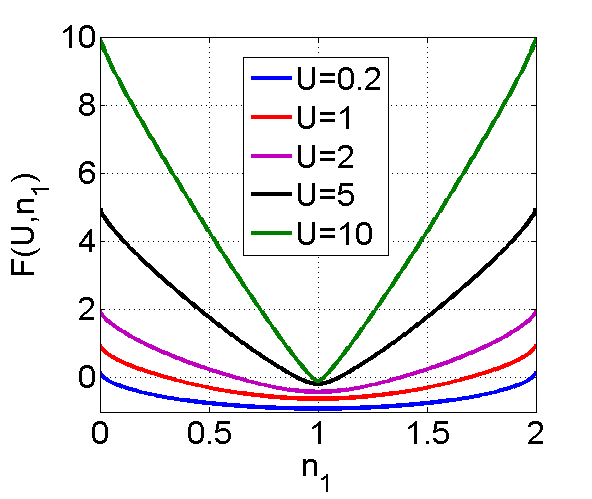}
\caption{
F-function(al) of Hubbard dimer as a function of $\no$ for several values of $U$ and $2\,t=1$.}
\label{ffunctional}
\end{figure}
We plot in Fig. \ref{ffunctional} the $F$-function(al) as a function of $\no$ for several values
of $U$.  
As $U$ increases we can see $F$ appears to tend to  $U |1-\no|$.

For any real problem the Euler equation for a given $\deps$ is
\ben
\label{euler}
\frac{d F(\no)}{d\no} -  \frac{\deps}{2} = 0,
\een
and the unique $\no(\deps)$ is found that satisfies this.  Then
\ben
E(\deps) = F(\no,\deps) - \deps\dn(\deps)/2.
\een
The oldest form of DFT (Thomas-Fermi theory\cite{T27,F28}) approximates
both $T(\no)$ and $V\ee(\no)$ and so leads to a crude treatment of the energetics of the system.
A variation on this was used in Ref. \cite{CC05} to enable extremely large calculations.

\ssec{Kohn-Sham method}
\label{KSmethod}

The modern world uses the KS scheme, and not pure DFT\cite{B12}.  The scheme in principle allows one to find the 
{\em exact} ground-state energy and density of an interacting problem by solving a 
non-interacting one.  This scheme is what produces such high accuracy while using simple
approximations in DFT calculations today.
Next, we see how the usual definitions of KS-DFT should be made for our dimer.

The heart of the KS method is the fictitious system of non-interacting electrons
whose density matches with the ground-state density of the interacting system.  
For our two-electron system, the KS system is that of non-interacting electrons ($U=0$)
with an on-site potential difference $\Delta v\s$, {\em defined} to reproduce the exact $\dn$  
of the real system.  This is just the tight-binding problem with an effective on-site potential
difference, and is illustrated in Fig. \ref{cartB}.

As stated in Section \ref{sec:dft}, in KS-DFT one conventionally extracts the Hartree contribution
from the electron-electron repulsion.  There are deep reasons for doing so,
which center on the remnant, the XC energy,
being amenable to local and semilocal-type approximations\cite{BPE98,PGB15}.
To see how the Hartree energy should be defined here, rewrite the
electron-electron repulsion as:
\ben
\hatV\ee = \frac{U}{2} \sum_i (\hat{n}_i^2 - \hat{n}_{i\uparrow}^2 - \hat{n}_{i\downarrow}^2).
\een
This form mimics the treatment in DFT.  The first term depends only on the total 
(i.e. spin-summed) density, akin to Hartree in real-space DFT.  The remaining terms 
cancel the self-interaction that arises from using the total density
for the electron-electron
interaction.  For the $N=2$ dimer, this decomposition results in   
\ben
U\H(\dn) = \frac{U}{2}\bUh,%\left(1 + (\dn/2)^2\right)
\een
and
\ben
E\x(\dn) = -\frac{U}{4} \bUh,%\left(1 + (\dn/2)^2\right)
\een
which satisfies $E\x = -U\H/2$ for $N=2$ as defined in real-space DFT for a spin singlet, Eq. (\ref{MFdef}).  
Together, the Hartree-Exchange is
\ben
\label{UHx}
E\Hx(\dn) = \frac{U}{4}\,\left(\no^2 + \nt^2\right) =\frac{U}{2} \left(1 + \left(\frac{\dn}{2} \right)^2\right).
\een
In Appendix \ref{limits} 
we see that the leading order in the $U$ expansion of the $F-$function(al) yields the same 
result. 
A typical mean field treatment of $\hat{V}\ee$ also results in Eq. (\ref{UHx}).  
In DFT there is always self-exchange, even for one or two particles. In
many-body theory, 
exchange means only exchange between different electrons.
Despite this semantic difference, both approaches yield the same leading-order-in-$U$
expression for the dimer, which we call $E\Hx$ here (but is often called just Hartree in
many-body theory).

For the dimer, from Eq. (\ref{tsdimer}), the KS kinetic energy is just 
\ben
\label{Ts}
T\s(\no)=-2\,t\Ts,
\een
so that $F\HF(\no) = T\s(\no) + E\Hx(\no)$ as in Section \ref{quanchem}. 
We can then define the correlation energy function from Eq. (\ref{dftcorr}),
so that
\ben\label{E_c_def}
E\c(\no)=F(\no)-T\s(\no)-E\Hx(\no).
\een%
\begin{figure}[htb] %
\includegraphics[width=1\columnwidth]{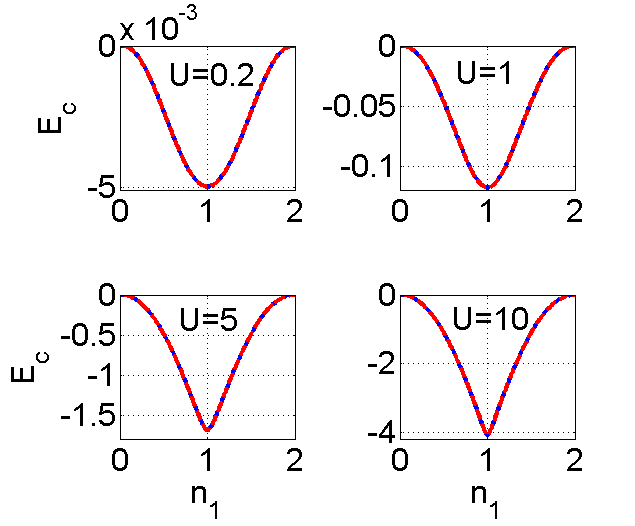}%
\caption{Plot of exact $E\c$ (blue line) and $E_{{\sss C},\rm{par}}$ (red dashed line) for different $U$ and $2\,t=1$.}
\label{Ec}%
\end{figure} %
In Fig. \ref{Ec}, we plot the correlation energy as a function of $\no$.
For small $U$, 
\ben
E\c \sim - U^2 (1-(\no-1)^2)^{5/2}/8 ~~~~~~ U\ll 2\,t
\een %(Eq. (\ref{Ecweak})) 
which is much smaller than the Hartree-exchange contribution, and is a relatively small contribution to $E$.  
But as $U$ increases, 
\ben
E\c\sim-U(1-(\no-1)^2)/2,~~~~~~~~ U\gg 2\,t
\een% (Eq. (\ref{Ecstrong})), 
with a cusp at half-filling.  Combined with
$E\Hx$, this creates $F$ for large $U$ as in Fig. \ref{ffunctional}.

Inserting this result into Eq. (\ref{euler}), we find that the KS electrons
have a non-interacting Hamiltonian:
\ben
\label{KSscA}
\hat{h}\s\, |\phi \rangle= \epsilon\s\, |\phi \rangle,
\een
where this KS Hamiltonian is
\ben
\label{KSscB}
\hat{h}\s(\dn) = -t\left(\hat{c}^{\dagger}_1\hat{c}_2 + h.c.\right)
+ \sum_i v_{s,i}(\dn) \hat{n}_i.
\een
The KS potential difference is
\ben
\label{KSpot}
\dvs(\dn) = \dv - U \dn/2 + \dv\c(\dn),
\een
where
\ben
\label{vcderiv}
\dv\c = -2\,dE\c(\no)/d\no.
\een
This is the key formal result of the KS paper\cite{KS65} as applied to SOFT:
  For any given
form of the (exchange-)correlation energy, differentiation yields the
corresponding KS potential.
If the exact expression for $E\c(\no)$ is used, 
this potential is guaranteed\cite{WSBW13} to yield the exact ground-state density
when the KS equations are iterated to convergence via a simple algorithm.

\begin{figure}[htb]
\includegraphics[width=\textwidth,height=0.7\textheight,keepaspectratio]{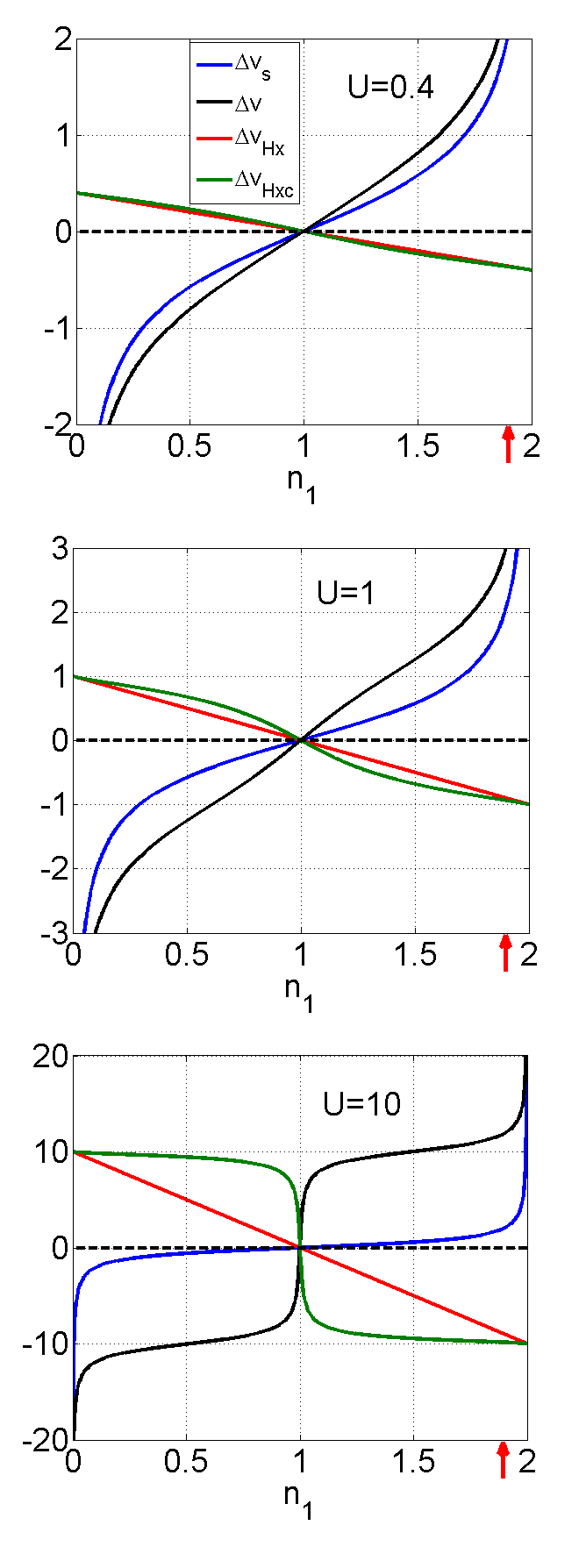}
\caption{Plots of $\Delta v\s$ (blue) and its components,
$\deps$ (black), $-U\dn/2$ (green), and $\Delta v\c - U\dn/2$ (red) plotted against $\no$ for
various $U$ and $2\,t=1$.
The arrows indicate the occupations used in Fig. \ref{cartB}.}
\label{dvc}
\end{figure}
In Fig. \ref{dvc}, we plot several examples of the dependence of the potentials
in the KS system as a function of $\no$, which range from weakly ($U=0.4$)
to strongly ($U=10$) correlated cases.  In each curve, the black line is
the actual on-site potential difference as a function of occupation of
the first site.  The blue line is the KS potential difference, which is
the on-site potential needed for two non-interacting ($U=0$) particles
to produce the given $n_1$.  This is found by inverting the tight-binding
equation for the density, Eq. (\ref{TBdv}).  Their difference is the Hartree-exchange-correlation
on-site potential, denoted by the red line.  Finally, the green line is just Hartree-exchange,
which ignores correlation effects.  For $U=0.4$, we see that the difference
between blue and black is quite small, and almost linear.  Indeed the Hartree-exchange
contribution is always linear (see Eq. (\ref{KSpot})).  Here the red is indistinguishable by eye
from the green, showing how small the correlation contribution to the potential is.
This means the HF and exact densities will be virtually (but not quite) identical.
When we increase $U$ to 1, we see a similar pattern, but now the red line is noticeably
distinct from the green.  
For any given $n_1$, the blue curve is smaller in magnitude than the black.  This is because
turning on $U$ pushes the two occupation numbers closer, and so their KS on-site potential
difference is smaller.  Again, the red curve is larger in magnitude than the green, showing
that HF does not suppress the density difference quite enough.   In our final panel, $U=10$,
and the effects of strong correlation are clear.  Now there is a huge difference between
black and blue curves.  Because $U$ is so strong, the density difference is close to zero for
most $n_1$, making the blue curve almost flat except at the edges.  In the KS scheme, this
is achieved by the red curve being almost flat, except for a sudden change of sign near $n_1=1$. 
These effects give rise to the $\dvs$ values shown in Fig. \ref{cartB}.
This effect is completely missed in HF.

\begin{figure}[htb] 
\includegraphics[width=1\columnwidth]{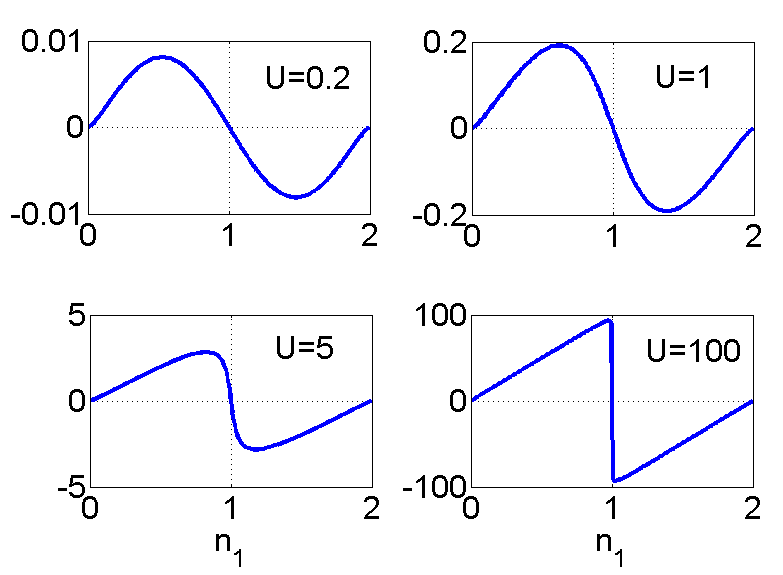}
\caption{Plot of $\Delta v\c$ for different $U$ and $2\,t = 1$.}
\label{vc}
\end{figure} 
To emphasize the role of correlation, in Fig. \ref{vc}, we plot the correlation potential alone,
which is the difference between the red and green curves in Fig. \ref{dvc}.  
Values from the blue curves for $\deps=2$ were used to make Fig. \ref{cartB}. 
$\Delta v\c$ is an odd 
function of $n_1$. In the weak- and strong-coupling limits we can write down simple expressions
for $\Delta v\c$ (see appendix \ref{correxp}): 
\bea
\Delta v\c &\approx& -\frac{5\,U^2 \dn}{32\,t}(1-(\dn/2)^2)^{3/2} ~~~~~~~~(U\ll 2\,t)\\
\Delta v\c &\approx& -U(1- |\dn/2|)\sgn(\dn)~~~~~~(U\gg 2\,t)
\eea
These correspond to the 1st and 4th panels in Fig. \ref{vc}.
For small $U$, it is of order $U^2$ (see appendix \ref{limits}), and has little effect.
As $U$ increases, it becomes proportional to $U$, and becomes almost linear in $U$, with a 
large step near $n_1=1$.  If we now compare this figure with Fig. \ref{dvc}, we see that it is
simply the derivative of 
the previous $E\c(\no)$ curve, as stated in Eq. (\ref{vcderiv}).

The self-consistent KS equations, Eqs. (\ref{KSscA}) and (\ref{KSscB}), have, in this case, precisely the same form as those
of restricted HF (or mean-field theory), Eqs. (\ref{MF_hamiltonian}) and (\ref{MFeff}), but with whatever additional dependence
on $\no$ occurs due to $\Delta v\c(\no)$.  
When converged, the ground-state energy is found simply from:
\ben
\label{E(n1)}
E(\no) = T\s(\no)+V\ext(\no)+U\H(\no) + E\xc(\no).
\een
The energy can alternatively be extracted from the KS orbital energy via Eq. (\ref{orbsum}):
\ben
\label{orbsum2}
E = 2\epsilon\s + (E\c + \Delta v\c \dn/2 - E\Hx),
\een
where the second term is the double-counting correction.
But note the crucial difference here.  We consider HF an approximate solution to the
many-body problem whereas DFT, with the exact correlation function(al), yields
the exact energy and on-site occupation, but not the exact wavefunction.

\sec{The fundamental gap}
\label{gap}

Now that we have carefully defined what exact KS DFT is for this
model, we immediately apply this knowledge to investigate a thorny
subject on the border of many-body theory and DFT, namely the fundamental
gap of a system.

\ssec{Background in real space}
\label{gapback}
Begin with the ionization energy of an $N$-electron system:
\ben
I = E(N-1)-E(N),
\een
is the energy required to remove one electron entirely from a system.
We can then define the electron affinity as the energy gained by adding
an electron to a system, which is also equal to the ionization energy
of the $(N+1)$-electron system:
\ben
A = E(N) - E(N+1).
\een
In real-space, $I$ and $A \geq 0$.
For systems which do not bind an additional electron, such as the He
atom, $A=0$.   The charge, or fundamental, gap of the system is then
\ben
E_g = I - A,
\een
and for many materials, $E_g$ can be used to decide if they are metals
($E_g=0$) or insulators ($E_g > 0$)\cite{K64}.
The spectral function of the single-particle Green's function has
a gap equal to $E_g$. For Coulombic matter, $E_g$ has always been 
found to be non negative, but no general proof has been given.

\def\ehomo{\epsilon\HOMO}
\def\elumo{\epsilon\LUMO}
Now we turn to the KS system of the $N$-electron system.
We denote the highest occupied (molecular) orbital as $\ehomo$
and the lowest unoccupied one as $\elumo$.
Then the DFT
version of Koopmans' theorem\cite{PPLB82,PL83,SS83,AP84,AB85,CVU10} shows that
\ben
\label{koop}
\epsilon\HOMO = - I,
\een
by matching the decay of the density away from any finite system in real
space, in the interacting and KS pictures.
However, this condition applies only to the HOMO, not
to any other occupied orbitals, or unoccupied ones.  In particular, the
LUMO level is not at $-A$, in general.
Define the KS gap as
\ben
E_{gs} = \elumo - \ehomo.
\een
Then $E_{gs}$ does not match
the true gap, even with the exact XC functional\cite{SP08,BGM13}.   We write
\ben
E_g = E_{gs} + \Delta\xc
\een
where $\Delta\xc \neq 0$, and is called the derivative discontinuity
contribution to the gap (for reasons that will be more apparent later)\cite{Pb85,Pb86}.
In general, $\Delta\xc$ appears to always be positive, i.e., the KS gap
is smaller than the true gap.  In semiconductors with especially small gaps,
such as germanium, approximate KS gaps are often zero, making the material a band
metal, but an insulator in reality.  The classic example of a chain of
H atoms becoming a Mott-Hubbard insulator when the bonds are stretched
is demonstrated unambiguously in Ref. \cite{SWWB12}.

While this mismatch occurs for all systems, it is
especially problematic for DFT calculations of
insulating solids.
For molecules, one can (and does) calculate the gap (called the
chemical hardness in molecular systems\cite{PY89}) by adding and removing
electrons.  But with periodic boundary conditions, there is no simple
way to do this for solids.
Even with the exact functional, the KS gap does not match the
true gap, and  there's no easy way to calculate $E_g$ in a periodic code.
In fact, popular approximations like LDA and GGA mostly produce good
approximations to the KS gap, but yield $\Delta\xc=0$ for solids.
Thus there is no easy way to extract a good approximation to
the true gap in such
DFT calculations.   The standard method for producing accurate gaps
for solids has long been to perform a GW calculation\cite{AG98}, an approximate calculation
of the Green's function, and read off its gap.  This works very well for
most weakly correlated materials\cite{SKF06}.  Such calculations are now done
in a variety of ways, but usually employ KS orbitals from an approximate
DFT calculation.  Recently, hybrid functionals like HSE06\cite{HSE06} have been
shown to yield accurate approximate gaps to many systems, but these
gaps are a mixture of the quasiparticle (i.e., fundamental) gap, and the
KS gap.  Their exchange component produces the fundamental gap at the HF level, which
is typically a significant overestimate, which then compensates for the
`too small' KS gap.  While this balance is unlikely to be accidental,
no general explanation has yet been given.

\ssec{Hubbard dimer gap}
\label{dimgap}

\begin{figure}[htb]
\includegraphics[width=.9\columnwidth]{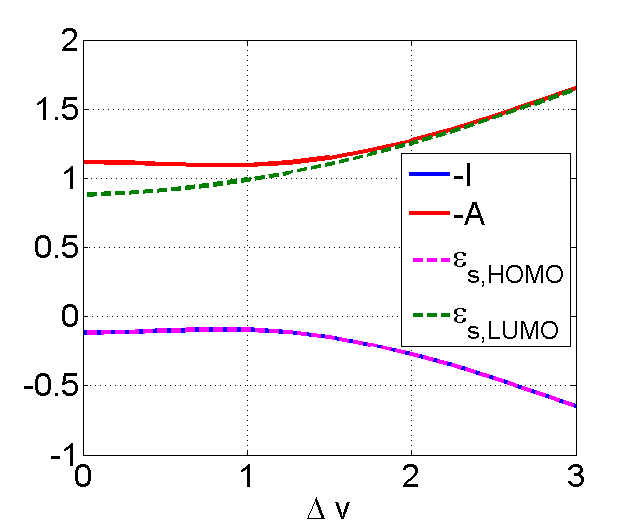}
\caption{
Plot of $-A$, $-I$, $\epsilon\HOMO$, and $\epsilon\LUMO$ as a function of $\deps$ with $U=1$ and $2\,t=1$.
}
\label{N2minusU1}
\end{figure}

\begin{figure}[htb]
\includegraphics[width=.9\columnwidth]{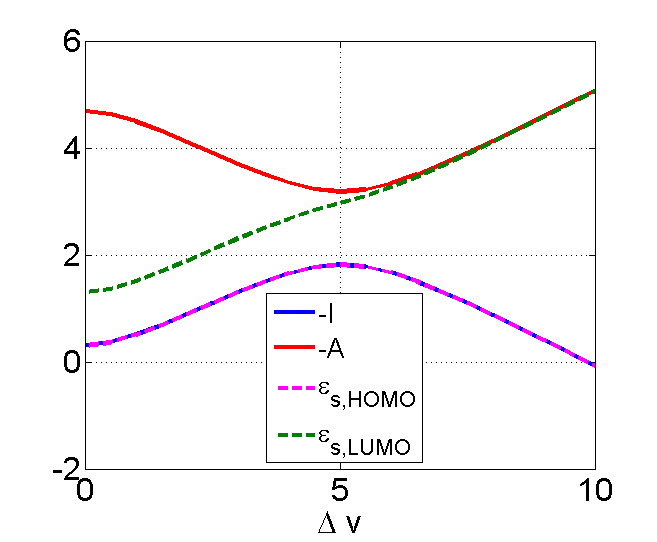}
\caption{
Plot of $-A$, $-I$, $\epsilon\HOMO$, and $\epsilon\LUMO$ as a function of $\deps$ with $U=5$ and $2\,t=1$.
}
\label{N2minusU5}
\end{figure}

For our half-filled Hubbard dimer, we can easily calculate both the $N\pm 1$-electron
energies, the former via particle-hole symmetry from the latter\cite{CFb12}.
In Fig. \ref{N2minusU1}, we plot $-I$, $-A$,
$\ehomo$, and $\elumo$ for $U=1$ when $2\,t=1$,
as a function of $\deps$.  We see that $A$ (and even sometimes $I$) can be negative
here.   (This cannot happen for real-space calculations, as electrons
can always escape to infinity, so a bound system always has $A \geq 0$.)
The HOMO level is always at $-I$ according to Eq. (\ref{koop})
but the LUMO is not at $-A$.  Here it is smaller than $-A$, and we find this
result for all values of $U$ and $\dv$.  The true gap is $I-A$, but the KS
gap is $\epsilon\LUMO+I$, which is always smaller.  Thus $\Delta\xc \geq 0$,
just as for real systems.

Fig. \ref{N2minusU1} is typical of weakly correlated systems, where $\Delta\xc$
is small but noticeable.  In Fig. \ref{N2minusU5}, we repeat the calculation with $U=5$, where
now $E_g \gg E_{gs}$ at $\dv=0$, but we still see the difference become tiny
when $\dv > U$.  In both figures, $\Delta\xc$ is the difference between the
red line and the green dashed line.  In all cases, $\Delta\xc \geq 0$, and
this has always been found to be true in real-space DFT, but has never been
proven in general.

\ssec{Green's functions}
\label{Green}

To end this section, we emphasize the difference between the KS and many-body
approaches to this problem by calculating their spectral functions\cite{ORR02}.
We define the many-body retarded single-particle Green's function as
\ben
\label{MBG}
G_{ij\sigma\sigma'}(t-t') = -i\, \theta(t-t') \langle \Psi_0 
|\{\hat{c}_{i\sigma}(t), \hat{c}^{\dagger}_{j\sigma'}(t') \}\,  |\Psi_0\rangle
\een
where $i,j$ label the site indices, $\sigma$, $\sigma'$ the electron spins, 
and $\{A,B\}=AB+BA$.
%\bea
%\label{Gdef}
%G_{ij}
%(\omega)&=&2\sum_\alpha\,\frac{|<\psi_\alpha^{N+1}|\,\hat{c}_{1\sigma}^{\dagger}\,|\psi_0^N>|^{2}}
%{\omega+E_{0}^N-E_\alpha^{N+1}+i\,\delta}\nonumber\\
%&&+\sum_\alpha\,
%\frac{|<\psi_\alpha^{N-1}|\,\hat{c}_{1\sigma}\,|\psi_\alpha^N>|^{2}}{\omega-E_0^N+E_\alpha^{N-1}+i\, \delta}
%\eea
For the Hubbard dimer at $N=1$ and $3$,  $|\Psi_0\rangle$ is a degenerate Kramers doublet and we      choose here the spin-$\uparrow$ partner.
Fourier transforming into frequency, we find for the diagonal component:
\bea
\label{Gdef}
G_\sigma(\omega)=G_{11\sigma\sigma}(\omega)
&=&\sum_\alpha\,\frac{|M^{\alpha}_{1\sigma}|^{2}}
{\omega+E^N-E_\alpha^{N+1}+i\,\delta}\nonumber\\
&&+\sum_\alpha\,
\frac{|L^{\alpha}_{1\sigma}|^{2}}{\omega-E^N+E_\alpha^{N-1}+i\,\delta}
\eea
where 
$M^{\alpha}_{1\sigma} = \langle\psi_\alpha^{N+1}|\,\hat{c}_{1\sigma}^{\dagger}\,|\psi_0^N \rangle$,
$L^{\alpha}_{1\sigma} = \langle\psi_\alpha^{N-1}|\,\hat{c}_{1\sigma}\,|\psi_0^N\rangle$, and
$\delta >0$ is infinitesimal.
Here, $\alpha$ runs over all states of the $N\pm1$-particle systems.
The other components have analogous expressions.
From any component of $G$, we find the corresponding spectral function
\ben
A(\omega)=- \Im G(\omega)/\pi
\een
We represent the spectral function $\delta$-function poles with lines whose height is proportional
to the weights.  Via a simple sum-rule\cite{FW71}, the sum of all weights in the spin-resolved        spectral function 
is $1$. There are four quasi-particle peaks for $N=2$. These peaks are reflection-symmetric
about $\omega=U/2$ for the symmetric dimer.

We also need to calculate the KS Green's function, $G\s(\omega)$.  This is done 
by simply taking the usual definition, Eq. (\ref{MBG}), and applying it to the ground-state
KS system.  This means two non-interacting electrons sitting in the
KS potential. The numerators vanish for all but single excitations. Thus the
energy differences in the denominators become simply occupied and unoccupied
orbital energies.  Since there are only two distinct levels (the  positive and
negative combinations of atomic orbitals), there are only two peaks, positioned
at the HOMO and LUMO levels, with weights:
\ben
M^{\alpha}_{1\sigma}=\frac{1}{2}\,
\left(1+\frac{\Delta v_s/2}{\sqrt{(\Delta v_s/2)^2+t^2}}\right), ~~~~ \rm{(KS)}
\een
and the sign between the contributions on the right is negative in the $L$ term.
Thus the symmetric dimer has KS weights of $1/2$.

\begin{figure}[htb]
\includegraphics[width=.9\columnwidth]{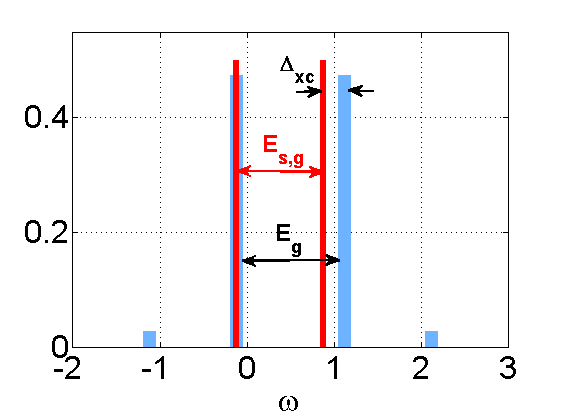}
\caption{
Spectral function of symmetric dimer for $U=1$, $\deps = 0$, and $2\,t=1$. 
The physical MB peaks are plotted in blue, the KS in red.
Here $I=0.1$,
$A=-1.1$, and $\epsilon\LUMO=0.9$, corresponding to $\dv = 0$ in Fig. \ref{N2minusU1}.}
\label{spectralsym}
\end{figure}
In Fig. \ref{spectralsym} we plot the spectral functions
for the symmetric case, for $U=1$, when $2\,t=1$.
Each pole contributes a delta function at a distinct transition frequency,
which is represented by a line whose height represents the weight.
The sum of all such weights adds to $1$ as it should, and the peaks are reflection-symmetric
about $U/2=0.5$.   The gap is the distance
between the highest negative pole (at $-I$) and the lowest positive pole 
(at $-A$).
We see
that the MB spectral function also has peaks that correspond to higher and lower 
quasi-particle excitations.  If we now compare this to the {\em exact} KS Green's function 
$G\s$, we see that, by construction, $G\s$ always has a peak at $-I$, whose weight need 
not match that of the MB function. It has only two peaks, the other being
at $\epsilon\LUMO$, which does not coincide with the position of the MB peak.
This is so because the KS scheme is defined to reproduce the ground-state occupations, nothing else.
But clearly, when $U$ is sufficiently small, it is a rough mimic of the MB
Green's function.  The larger peaks in the MB spectral function each have KS
analogs, with roughly the correct weights.  One of them is even at exactly
the right position.   Thus if a system is weakly correlated, the KS spectral
function can be a rough guide to the true quasiparticle spectrum.

\begin{figure}[htb]
\includegraphics[width=.9\columnwidth]{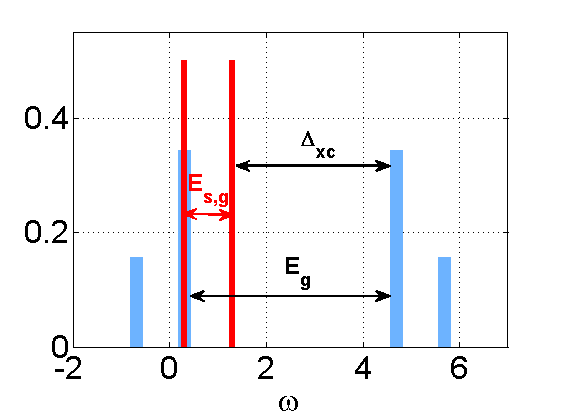}
\caption{Same as Fig. \ref{spectralsym}, but now $U=5$. 
Here $I=-0.3$, $A=-4.7$, and $\epsilon\LUMO=1.3$, corresponding to $\dv = 0$ in Fig. \ref{N2minusU5}.
}
\label{bigU}
\end{figure}
On the other hand, when $U \gg 2\,t$, the KS spectral function is
not even close to the true MB spectral function,  as illustrated in Fig. \ref{bigU}.  
Now the two lowest-lying MB peaks approach each other, as do the two highest lying peaks,
therefore increasing the quasi-particle gap. In addition, the weights tend to equilibrate with
each other. In fact, when $U\to\infty$ and/or $t\to 0$, those two lowest-lying peaks gather
together at $\omega = 0$, having both the same weight of $1/4$. And
similarly the two highest-lying peaks merge at $\omega=U$, also with 
a weight of $1/4$. They become the precursors of the lower and upper 
Hubbard bands  with a quasi-particle gap equal to $U$. If more sites are added to the symmetric
dimer, other quasi-particle peaks appear, that also merge into the lower and upper Hubbard
bands as $U\to\infty$. Notice that the spectral function
has significant weights for transitions between states that
differ from the HOMO and LUMO, and are forbidden in the KS spectral
function for large $U$.   
In Fig. \ref{bigU}, we see that not only there is a large difference
between the gaps in the two spectral functions, but also the KS weights are
not close to the MB weights.  The only `right' thing about the KS
spectrum is the position of the HOMO peak.

\begin{figure}[htb]
\includegraphics[width=.9\columnwidth]{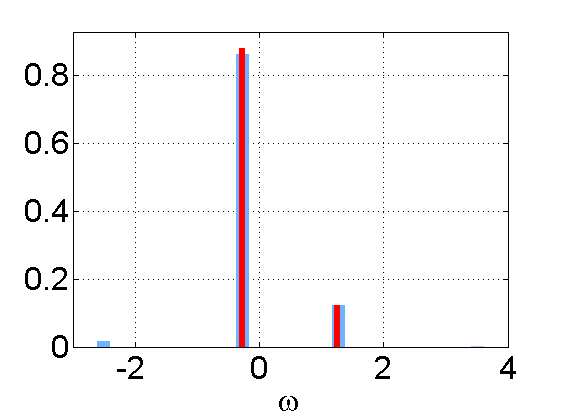}
\caption{Same as Fig. \ref{spectralsym}, but now  $U=1$, $\dv=2$.
Here $I=0.27$, $A=-1.27$, and $\epsilon\LUMO =1.25$, 
corresponding to $\dv =2$ in Fig. \ref{N2minusU1}.
}
\label{spectralasym}
\end{figure}
In Fig. \ref{spectralasym}, we plot the spectral functions for $\deps=2$ and $U=1$, to see the effects
of asymmetry on the spectral function.  Now the system appears entirely uncorrelated, and the KS spectral function
is very close to the true one, much more so than in the symmetric case.  
Here $\Delta\xc$ is negligible. 
The asymmetry of the potential strongly suppresses correlation effects.   
\begin{figure}[htb]
\includegraphics[width=.9\columnwidth]{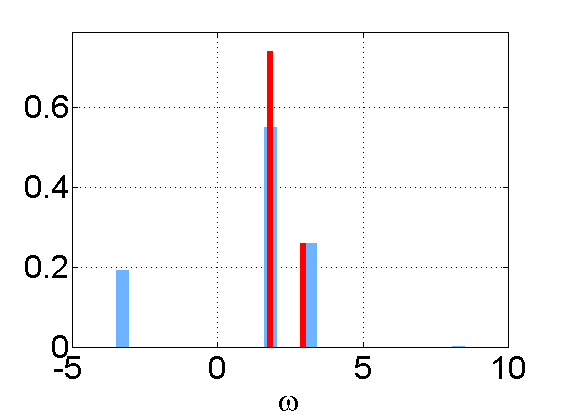}
\caption{Same as Fig. \ref{spectralasym}, but now $U=5$, $\dv=5$.
Here $I=-1.8$, $A=-3.2$, and $\epsilon\LUMO=3$, 
corresponding to $\dv =5$ in Fig. \ref{N2minusU5}.
}
\label{bigUbigv}
\end{figure}
In Fig. \ref{bigUbigv}, we see that the effects of strong $U$ are
largely quenched by a comparable $\dv$.  Here $\Delta\xc$ is small
compared to the gap, but not all KS peak heights are close to their
MB counterparts.

The situation is interesting even for the `simple' case, $N=1$, in which
the ground-state is open-shell\cite{GBc04}. Here the interacting 
spin-$\uparrow$ and -$\downarrow$ Green's functions differ.
To understand why, we 
choose the $N=1$ ground state to have spin $\uparrow$. This state has energy 
$E(1)=-\sqrt{t^2+(\dv/2)^2}$. Adding a $\downarrow$-spin electron takes the system
to the different singlet states at $N=2$, and to the triplet state with $S_z=0$.
One of them is the ground state at $N=2$ whose energy $E(2)<0$ is given in Eq. (\ref{eq:gsenergy})
in the appendix. In contrast, adding an $\uparrow$-spin electron takes the interacting 
system to the triplet $N=2$ state with $S_z=1$, whose energy is trivially given by 
$E(2)_\mathrm{trip}=0$. Annihilating an $\uparrow$-spin electron takes the system
to the vacuum, while it is impossible to annihilate a $\downarrow$-spin electron. 
These clearly illustrates that the number and energy of the poles in 
$G_{\uparrow}$ and $G_{\downarrow}$ is different: 
$G_{\uparrow}$ has only two quasi-particle peaks, with trivial energies
$E(2)_\mathrm{trip}-E(1)=\sqrt{t^2+(\dv/2)^2}$ and $E(1)-E(0)=-\sqrt{t^2+(\dv/2)^2}$.
This last expression corresponds to the ionization energy $I=E(0)-E(1)=\sqrt{t^2+(\dv/2)^2}$.
$G_{\downarrow}$ has four quasiparticle peaks, all corresponding to adding
a $\downarrow$-spin electron, with non-trivial energies. The lowest of these corresponds
to the electron affinity $A=E(1)-E(2)=-\sqrt{t^2+(\dv/2)^2}-E(2)$.
In other words, ionization involves either removing an $\uparrow$-spin electron (hence seen
as a pole in $G_{\uparrow}$) or adding a $\downarrow$-spin electron (hence seen as
a pole in $G_{\downarrow}$). The interacting gap is $E_g=I-A=2\,\sqrt{t^2+(\dv/2)^2}+E(2)$.

We turn now to the KS Green's function. For $N=1$, the KS on-site potentials equal the 
true on-site potentials, $\pm\dv/2$. So the ground-state (chosen again to have spin $\uparrow$)
has energy $E\s(1)=-\sqrt{t^2+(\dv_s/2)^2}$. Since the other state has energy $E\s(1)$, and
a second $\up$-electron occupies that state,
the total KS energy
is $E(2)_\mathrm{S_z=1}=0$.  On the other hand, annihilating the $\uparrow$ electron
costs an energy $E(1)$. This shows that the $\uparrow$-spin KS and interacting
Green's functions are identical to one other and trivial for $N=1$. Thus $I=-\epsilon\HOMO=\sqrt{t^2+(\dv/2)^2}$.
This result is specific to this model.

Removing a $\downarrow$-spin KS electron is impossible, just as in the interacting case. However, adding it means
having either two opposite-spin KS electrons with the same energy $-\sqrt{t^2+(\dv_s/2)^2}$, or having
one with energy $-\sqrt{t^2+(\dv_s/2)^2}$ and another with energy $\sqrt{t^2+(\dv_s/2)^2}$. The first case
corresponds to the KS ground-state with energy $-2\,\sqrt{t^2+(\dv_s/2)^2}$, while the second one
is an excited state with energy 0. The KS value for the electron affinity is $A_s=E\s(1)-E\s(2)=\sqrt{t^2+(\dv_s/2)^2}$,
which differs from the interacting value. Furthermore, the KS gap $E_{gs}=0$ is clearly 
an incorrect estimate of the true interacting gap, which is given by $I=\Delta_{xc}$.

\begin{figure}[htb]
\includegraphics[width=.9\columnwidth]{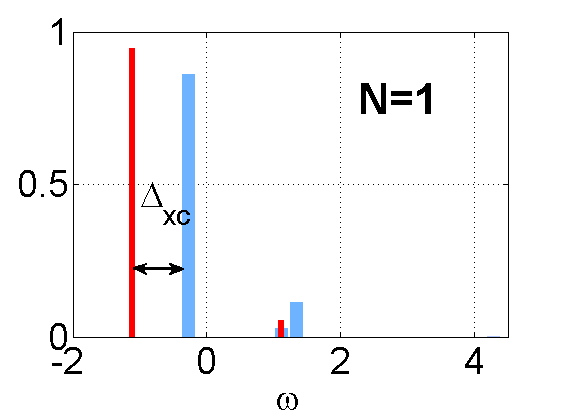}
\caption{Spin-$\downarrow$ resolved spectral function for $N=1$ and $U=1$, $\dv=2$. 
Here $I=1.12$, $A=0.27$, and $\epsilon\LUMO=\epsilon\HOMO=-1.12$.}
\label{GFdv2U1N1}
\end{figure}

\begin{figure}[htb]
\includegraphics[width=.9\columnwidth]{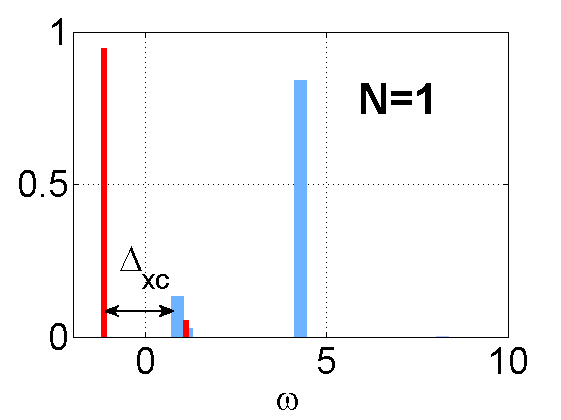}
\caption{Spin-$\downarrow$ resolved spectral function for $N=1$ and $U=5$, $\dv=2$.
Here $I=1.12$, $A=-0.90$, and $\epsilon\LUMO=\epsilon\HOMO=-1.12$.}
\label{GFdv2U5N1}
\end{figure}

Figs. \ref{GFdv2U1N1} and \ref{GFdv2U5N1}  show the spectral function associated
with $G_{\downarrow}$ for the
many-body and KS Green's functions for $N=1$ and $\dv=2$.  In the first,
$U=1$, so it is relatively asymmetric, whereas in the second, $U=5$, making
it close to symmetric. 
Thus the HOMO is at the lowest red line, and matches exactly the LUMO, with
a KS gap of zero.  Thus $\Delta\xc$ {\em is} the gap of the interacting system.
We see that in the first figure, correlations are weak and the KS spectral function
mimics the physical one, but in the second figure ($U=5$), they differ substantially,
even though $N=1$!

The difference in expressions for spin species is illustrated further by work analyzing
Koopmans' and Janak's theorems for open-shell systems\cite{GB02,GBB03,GGB03,GBc04}.
Self-energy approximations beyond GW have been performed on the Hubbard dimer\cite{RGR09,RBR12}, 
as well as a battery of many-body perturbation theory methods\cite{OT14}
though only for the symmetric case.

The bottom line message of this subsection is that the KS spectral function does not match the
quasiparticle spectral function, because it is not supposed to.   However,
the main features of a weakly correlated system are loosely approximated
by those of the KS function, with the gap error shifting the upper part of
the spectrum relative to the lower part.  This is the motivation behind the
infamous scissors operator in solid-state physics.
A very accurate DFT approximation can (at best)
approximate the KS spectral function, not the many-body one.   The exact XC
functional does not reproduce the quasiparticle gap of the system.
For strongly correlated systems, there are often substantial qualitative differences
between the MB and KS spectral functions.  These are some of the limitations of KS-DFT.
that, e.g., DMFT is designed to overcome \cite{GKKR96}.

\sec{Correlation}
\label{Correlation}
\ssec{Classifying correlation: Strong, weak, dynamic, static, kinetic, and potential}

There are as many different ways to distinguish weak from strong correlation
as there are communities that study electronic structure.   Due to the limited
degrees of freedom (namely, one), these all overlap in the Hubbard dimer.  We
will discuss each.

The most important thing to realize is that correlation energy comes in two
distinct contributions:  kinetic and potential.   These are entirely well-defined
quantities within KS-DFT.   The kinetic correlation energy is:
\ben
\label{kincorr}
T\c = T - T\s
\een
for a given density.  Note that we could as easily call this the correlation
contribution to the kinetic energy.  The potential correlation energy is:
\ben
\label{potcorr}
U\c = V\ee -  E\Hx,
\een
and could also be called the correlation contribution to potential energy.
For future notational convenience,  we also define $U\x=E\x$, i.e., there is
no kinetic contribution to exchange.
Then, from Eq. (\ref{dftcorr}), we see 
\ben
E\c=T\c+U\c.
\een

\begin{figure}[htb] 
\includegraphics[width=1\columnwidth]{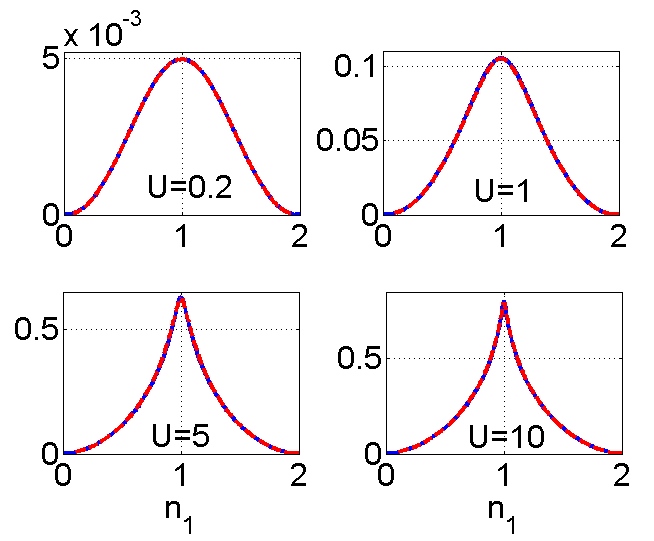}
\caption{Plot of exact $T\c$ (blue line) and $T_{{\sss C},\rm{par}}$ (red dashed line) for different $U$ and $2\,t=1$.}
\label{Tc}
\end{figure}

\begin{figure}[htb] 
\includegraphics[width=.9\columnwidth]{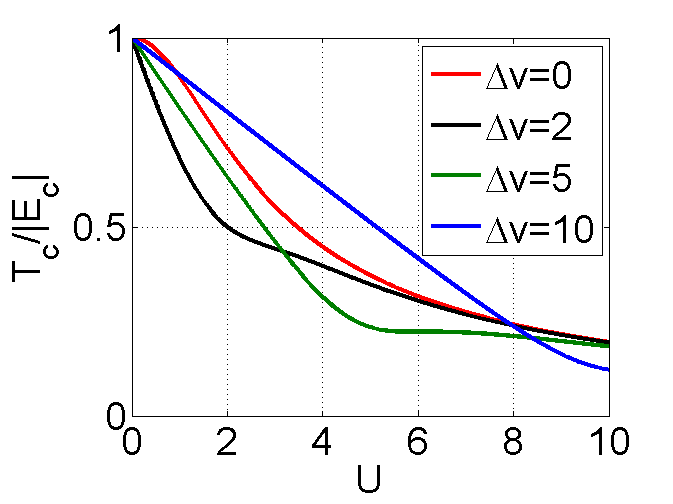}
\caption{Plot of $\beta_{\rm{corr}} = T\c/|E\c|$ as a function of $U$ with $2\,t=1$.}
\label{betacorr}
\end{figure}

We can now use these to discuss the differences between weak and strong correlation.
First note that, by construction, and as shown for our dimer in appendix \ref{proofs},
\ben
\label{corrineq}
E\c < 0,~~~T\c > 0,~~~~U\c < 0.
\een
In Figs. \ref{Ec} and \ref{Tc}, we plot both $E\c$ and $T\c$, respectively, for several values of $U$ (with $2\,t=1$).
When $U$ is small, $T\c \approx -E\c$.  However, for $U \gg 2\,t$, we see that
although $E\c$ becomes very large (in magnitude), $T\c$ remains finite and in fact,
$T\c$ never exceeds $2\,t$ as proven in Appendix \ref{proofs}.   We can define a measure of the nature of the correlation\cite{BEP97}:
\ben
\beta_{\rm{corr}} \equiv \frac{T\c}{|E\c|},
\een
As $U \to 0$, $\beta_{\rm{corr}} \to 1$, while as $U \to \infty$, $\beta_{\rm{corr}} \to 0$.
Thus $\beta_{\rm{corr}}$ close to 1 indicates weak correlation, $\beta$ small indicates
strong correlation.   We plot $\beta_{\rm{corr}}$ as a function of $U$ 
for several values of $\deps$ in Fig. \ref{betacorr}.
Although $\beta_{\rm{corr}}$ is monotonically decreasing with $U$ for $\dv =0$, we see that the
issue is much more complicated once we include asymmetry.  The curve for each $\dv$ remains
monotonically decreasing with $U$.  But consider $U=2$ and different values of $\dv$.  Then
$\beta_{\rm{corr}}$ at first decreases with $\dv$, i.e. becoming \emph{more} strongly correlated,
but then increases again for $\dv > U$, ultimately appearing less correlated than $\dv =0$.

Quantum chemists often refer to {\em dynamic} versus {\em static} correlation.
Our precise prescription in KS-DFT loosely corresponds to their definition, replacing
dynamic by kinetic, and static by potential.  Thus, considering an H$_2$ molecule
with a stretched bond, the Hubbard model applies.  As the bond stretches, $t$ vanishes,
and $U/2\,t$ grows.   Thus $\beta_{\rm{corr}} \to 0$ as $R \to \infty$.  The exact wavefunction,
the Heitler-London wavefunction\cite{HL35}, has only static correlation in this limit.  In many-body
language, it is strongly correlated.  In DFT language, the fraction of correlation
energy that is kinetic is vanishing.

\ssec{Adiabatic connection}
\label{sec:AC}

With the various contributions to correlation well-defined, we construct the adiabatic connection (AC) formula 
\cite{LP75,GL76} for the Hubbard dimer.
The adiabatic connection has had enormous impact on the field of DFT as it allows
both construction \cite{B93,Bb93,PEB96,ES99,AB99,MCY06}, and understanding \cite{PEB96,BEP97,PMTT08}, of exact and approximate functionals solely
from their potential contributions. 

In many-body theory, one often introduces a coupling-constant in front of the interaction.
In KS-DFT, a coupling constant $\lambda$ is introduced in front of the electron-electron
repulsion but, contrary to traditional many-body approaches, the density is held fixed
as $\lambda$ is varied (usually from 0 to 1). 
Via the Hohenberg-Kohn theorem, as long
as there is more than 1 electron, this implies that the one-body potential must
vary with $\lambda$, becoming $v\l(\br)$.  By virtue of the density being held fixed,
$v^{\lambda =0}(\br)=v\s(\br)$ while $v^{\lambda=1}(\br)=v(\br)$.  Thus $\lambda$ interpolates between the
KS system and the true many-body system.  Additionally, $\lambda\to\infty$
results in the strictly correlated electron limit\cite{SPL99,SGS07,LB09,GS10,MG12} which provides useful information about real systems that are strongly correlated.

\begin{figure}[htb]
\includegraphics[width=1\columnwidth]{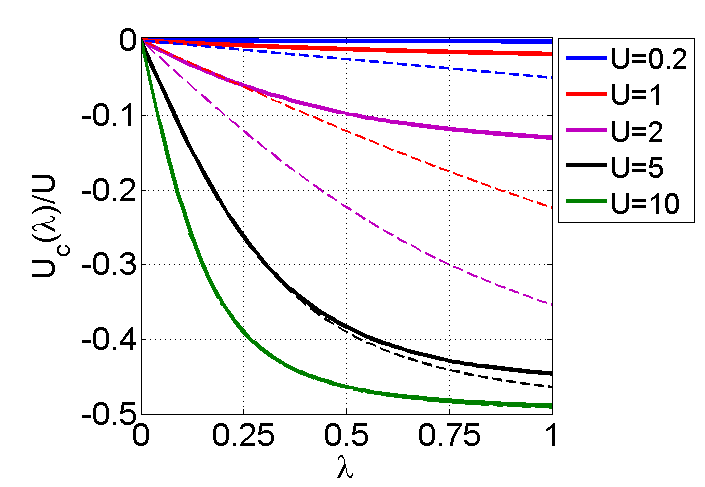}
\caption{Adiabatic connection integrand divided by $U$ 
for various values of $U$.  The solid lines are
$\deps =2$ and the dashed lines $\deps =0$.  Asymmetry reduces the correlation energy but increases the fraction of kinetic correlation. 
}
\label{ACplot}
\end{figure}

The adiabatic connection for the Hubbard dimer is very simple.  Define the XC energy
at coupling constant $\lambda$ by simply multiplying $U$ by $\lambda$ while keeping $\dn$ fixed:
\ben
E\xc^\lambda(U,\dn)=E\xc(\lambda U,\dn).
\een
Application of the Hellman-Feynman theorem\cite{F39} yields\cite{HJ74,LP75,LP77,GL76}:
\ben
\label{uclambda}
\frac{dE\xc(\lambda U,\dn)}{d\lambda} =  \frac{U\xc(\lambda U,\dn)}{\lambda},
\een
where $U\xc( U,\dn)$ is the potential contribution to the
XC energy, i.e., $U\x=E\x$ and
\ben
U\c(\lambda U) =V\ee (\lambda U) - \lambda\, E\Hx(U).
\een
Thus, we can extract $T\c$ solely from our knowledge of $E\c(U)$ via
\ben
\label{tclambda}
T\c =E\c - U\c= E\c - \left. \frac{d E\c\l}{d \lambda}\right|_{\lambda=1}.
\een
Thus, any formula for $E\c$, be it exact or approximate, yields a corresponding
result for $T\c$ and $U\c$, and vice versa\cite{FTB00}.
We may then write
\ben
\label{AC}
E\xc(U,\dn) = \int_0^1 \frac{d\lambda}{\lambda}\, U\xc(\lambda U,\dn),
\een
and this is the infamous adiabatic connection formula of DFT\cite{LP75,GL76}.
We denote the integrand as $U\c(\lambda)$, defined as
\ben
U\c(\lambda) = \frac{U\c(\lambda U)}{\lambda} = \frac{d E\c(\lambda U)}{d\lambda}.
\label{Ucldef}
\een
Plots of $U\c(\lambda)$ from  Eq. (\ref{Ucldef})
are called adiabatic connection plots, and can be used
to better understand both approximate and exact functionals.  
In Fig. \ref{ACplot}, we plot a typical case for $U=2\,t$ and $\dv=0$.  
They have the nice interpretation that the value at $\lambda=1$
is the potential correlation energy, $U\c$, the area under the
curve is $E\c$, and 
the area between the curve and the horizontal line at $U\c(1)$ is $-T\c$.
Furthermore, one can also show\cite{LP85}
\ben
\frac{d U\xc(\lambda)}{d\lambda} < 0,
\een
from known inequalities for $T\c(\lambda)$ and $E\c(\lambda)$.  
This is proven for our problem in appendix \ref{proofs}. 
Interestingly, such curves have always been found to be convex when extracted numerically
for various systems\cite{PCH01,FNGB05}, but no general proof of this is known.
The Hubbard dimer
also exhibits this behavior.  A proof for the dimer might suggest a proof for
real-space DFT.

In Fig. \ref{ACplot} we plot $U\c(\lambda)/U$
for $\dv=0$ and $\dv=2$, with various values of $U$.  
From the above formulas, one can deduce that the area between the curve and the horizontal
line at $U\c(1)$ is $-T\c$.
Thus as $U$ grows, the curve moves from being almost linear to decaying very
rapidly, and $\beta$ varies from 1 down to 0.  

In Fig. \ref{ACplot}, we show $U$ up to 10 (for $2\,t=1$), to show the effect of 
stronger correlation.  Not only has the magnitude of the correlation become
larger, but the curve drops more rapidly toward its value at large $\lambda$.
$\beta_{\rm{corr}} \simeq 0.9$ for $\Delta v=0$ and $U=1$, but $\beta_{\rm{corr}}  \simeq 0.2$ for $\Delta v=0$
and $U=10$, reflecting the fact that the increase in correlation is of the static kind.

The weakly correlated limit has been much studied in DFT.
Perturbation theory in the coupling constant is called Goerling-Levy 
perturbation theory\cite{GL93}.
For small $\lambda$, 
\ben
U\c(\lambda U) = \lambda^2 U\c^{(2)} + \lambda^3 U\c^{(3)}+...~~~(\lambda\to 0).
\een
In Appendix \ref{correxp}, we show that 
\ben
U\c^{(2)}(\dn)= -\frac{U^2}{8\,t} \left(1-\left(\frac{\dn}{2}\right)^2\right)^{5/2},
\een
and
\ben
U\c^{(3)} (\dn)= \frac{3\,U^3}{32\,t^2} \left(\frac{\dn}{2}\right)^2 \left(1-\left(\frac{\dn}{2}\right)^2\right)^3
\een
for the dimer.
This yields, for $T\c$,
%\ben
%T\c = -\frac{1}{2}U\c^{(1)}\lambda^2 -\frac{2}{3} U\c^{(2)}\lambda^3 %eq 37 in GL93 
%- \frac{3}{4} U\c^{(3)}\lambda^4 ...
%\een
\ben
T\c = -\frac{1}{2}\lambda^2 U\c^{(2)} - \frac{2}{3} \lambda^3 U\c^{(3)} - \frac{3}{4} \lambda^4 U\c^{(4)} - ...%modified eq 37 in GL93
\een
showing that $\beta \to 1$ as $U$ (or $\lambda$) vanishes.
For any system, $U\c^{(2)}$ determines the initial slope of $U\c(\lambda)$.  

On the other hand, in the strongly correlated limit, in real-space\cite{LB09,GSV09}.
\ben
E\c \to \lambda(B_0+ \lambda^{-1/2} B_1 + \lambda^{-1}B_2 ...),~~~~(\lambda\to \infty)
\een
where $B_k$ $(k=0,1,2...)$ are coupling-invariant functionals of $\n(\bm r)$\cite{LBb09}.
The dominant term is linear in $U$.  Physically, it must exactly cancel the 
Hartree plus exchange contributions, since there is no electron-electron
repulsion to this order when each electron is localized to separate sites.  
Correctly,
such a term cancels out of $T\c$, so that its dominant contribution is $O(1)$.  
From Appendix \ref{correxp}, we see that the Hubbard dimer has a different form, involving
only integer powers of $\lambda$:
\ben
E\c \to \lambda\, B_0+  \tilde B_1 +  \tilde B_2/\lambda +...~~~~(\lambda\to \infty)
\een
where
\ben
B_0(\dn) = -U (1-\dn/2)^{2}/2,
\een
\ben
\tilde B_1(\dn) =2\,t \sqrt{1-\dn/2}\, (\sqrt{1+\dn/2}-\sqrt{\dn}),
\een
and
\ben
\tilde B_2(\dn) = (1-\dn/2)t^2/U.
\een
But both this term and the next cancel in the total energy 
(at half filling), so that the ground-state energy is $O(1/U)$, i.e., extremely
small as $U$ grows:
\ben
E \to -\frac{4 t^2}{U}
\een
This illustrates that, although the KS description is exact, it becomes quite
contorted in the large $U$ limit.  This has been implicated in convergence
difficulties of the KS equations, even with the exact XC functional,  because
the KS system behaves so differently from the physical system\cite{WBSB14}.

\sec{Accurate parametrization of correlation energy}
\label{param}

Although the Hubbard dimer has an exact analytic solution when constructed from many-body
theory, the dependence of $F(\dn)$ (or equivalently $E\c(\dn)$) is only given implicitly.
While this is technically straightforward to deal with, in practice it would be much simpler
to use if an explicit formula is available.  In this section, we show how the standard
machinery of DFT can be applied 
to develop an extremely accurate parametrization of the correlation energy functional. 

An arbitrary antisymmetric wavefunction 
is characterized by 3 real numbers where $|12\rangle$ means an electron at site 1 and site 2, etc.:
\ben
\label{wf}
|\psi\rangle=\alpha\,\left(|12\rangle +|21\rangle \right)\,+\beta_1\,|11\rangle \,+\beta_2\,|22\rangle.
\een
Normalization requires $2\alpha^2 + \beta_1^2 +\beta_2^2 =1$. 
In terms of these parameters, the individual components of the
energy are rather simple:
\bea
T &=& -4\,t\,\alpha(\beta_1+\beta_2)\nonumber\\
V\ee&=& U(\beta_1^2+\beta_2^2)\nonumber\\
V&=&-\dv(\beta_1^2-\beta_2^2),
\label{comp}
\eea
so that the variational principle may be written as
\ben
E = \min_{\substack{\alpha, \beta_1, \beta_2\\ 1 
= 2\alpha^2 + \beta_1^2 + \beta_2^2}} E(\alpha,      \beta_1, \beta_2).
\een
The specific values of these parameters for the ground-state wavefunction are reported
in appendix \ref{energycomp}.

For this simple problem, we are fortunate that we can apply the Levy-Lieb
constrained search method explicitly.  A variation of this method was used for the derivation of the exact functional of the single- and double-site Anderson model and the symmetric Hubbard dimer\cite{CFb12}. The functional $F[\n]$ is defined
by minimizing the expectation value of $\hat T + \hat V\ee$ over all possible
wavefunctions yielding a given $\n(\br)$.  In real-space DFT, there are no
easy ways of generating interacting wavefunctions for a given density.  But here,
\ben
\dn = 2 (\beta_1^2 - \beta_2^2),
\een
which allows us to simply eliminate a parameter, e.g., $\beta_2$ in favor of $\dn$.  Thus 
\ben
\label{levy1lat}
F[\dn] = \min_{\alpha^2 + \beta_1^2 =\frac{1}{2} (1 + \frac{|\dn|}{2})} \left[T(\alpha,\beta_1,\dn) +V\ee(\alpha,\beta_1,\dn) \right].
\een
With normalization and the density constraint, only one parameter is left free.
There exist several possible
choices for this. If we choose $g=2\alpha\,(\beta_1+\beta_2)$ which corresponds to
the hopping term, then after some algebra the function(al) can be written nicely as
\ben
F(\rho)=\min_g\,f(\rho, g)
\een
with the intermediate quantity
\begin{equation}
\label{Fung}
f(\rho,g)=-2\,t\,g+U h(g,\rho),
\end{equation}
and
\ben
\label{h}
h(g,\rho) = \frac{g^2\,(1-\sqrt{1-g^2-\rho^2})+2\rho^2}{2(g^2+\rho^2)}.
\een
Note that both $t$ and $U$ appear linearly in $f(g,\rho)$. 
The minimization yields a sextic polynomial, equation (\ref{polynomialg}), that $g$ must satisfy.
The weak-coupling, strong-coupling, symmetric, and asymmetric limits of $g$ are given in 
appendix \ref{limits}.%, and an alternative decomposition for $F$ is in appendix \ref{limits2}.

Our construction begins with a simple approximation to $g(\rho)$:
\ben
\label{g0}
g_0(\rho)=\sqrt{\frac{(1-\rho)\,(1+\rho\,(1+(1+\rho)^3 u a_1(\rho,u)))}{1+(1+\rho)^3 u a_2(\rho,u)}}
\een
where
\ben
a_i(\rho,u) = a_{i1}(\rho) + u\, a_{i2}(\rho),
\een
and
\bea
a_{21}&=&\frac{1}{2}\sqrt{(1-\rho)\rho/2},~~~a_{11}=a_{21}(1+\rho^{-1}),       \nonumber\\
a_{12}&=&\frac{1}{2}\,(1-\rho),~~~~~~~~~~a_{22}=a_{12}/2.
\eea
These forms are chosen
so $g_0$ is exact to second- and first- order in the
weak- and strong-coupling limits respectively, and to first- and second- order in
the symmetric and asymmetric limits respectively.  Use of this $g_0$ to construct
an approximation to $F$, $f(g_0(\rho),\rho)$, yields very accurate energetics.
The maximum energy error, divided by $U$, is 0.002.

But for some of the purposes in this paper, such as calculations of $T\c$, even
this level of error is unacceptable.  We now improve on $g_0(\rho)$ using the
adiabatic connection formula of Sec \ref{sec:AC}.   Like $F$, we can define functions of two variables for
each of the correlation components.  Write
\ben
\label{Ec1}
e\c(g,\rho)=f(g,\rho) - T\s(\rho) -E\Hx(\rho).
\een
where $T\s$ and $E\Hx$ are from Eqs. (\ref{Ts}) and (\ref{UHx}), respectively.
The kinetic and the potential correlation are given by
\bea
\label{Tc1}
t\c(g,\rho)&=&T-T\s= -2\,t \left(g-\sqrt{1-\rho^2}\right)\\
\label{Uc1}
u\c(g,\rho)&=&V_{ee}-E\Hx
=U\left[h(g,\rho) - (1+\rho^2)/2 \right],
\eea
and their sum yields $e\c(g,\rho)$.
If we insert $g(\rho)$, the exact minimizer of $f(g,\rho)$, into any of these
expressions, we get the exact answers.

But recall also that one can extract $U\c$ from the derivative of $E\c$ with
respect to the coupling constant $\lambda$, i.e.,
\ben
\label{Uc}
U\c = dE\c(\lambda)/d\lambda|_{\lambda=1}.
\een
Now for any $g$ and $e\c(g)$, we can find the $\lambda$ dependence by replacing
$U$ by $\lambda U$.  Thus
\ben
\label{partialEc}
\frac{d e\c(g,\lambda)}{d \lambda}=
\frac{\partial E\c(\lambda)}{\partial \lambda}+
\frac{\partial E\c(\lambda)}{\partial g}\,\frac{\partial g}{\partial \lambda}
\een
Since $T\s$ and $E\Hx$ do not depend on $g$, the minimization of $f$ reduces to
$\partial e\c/\partial g =0$, so for the exact $g$ the second term on the right
of Eq. (\ref{partialEc})
is always zero.   But it does not vanish for $g_0$.

Equating Eqs. (\ref{Uc}) and (\ref{partialEc}) and using the definitions, we find the following
self-consistent equation for $g$:
\ben
g = -\frac{T}{2\,t} +\frac{1}{2\,t} \frac{\partial E\c}{\partial g} \left.
\frac{\partial g(\lambda)}{\partial\lambda}\right|_{\lambda=1}.
\een
We may use this to improve our estimate for $g$.  Simply evaluate the right-hand side
at $g_0$, to find:
\ben
\label{gtilde}
g_1 = g_( u \frac{\partial h}{\partial g}-1)\,
\left.\frac{\partial g(\lambda)}{\partial \lambda}\right|_{\lambda=1},
\een
where
\bea
\left.\frac{\partial g(\lambda)}{\partial \lambda}\right|_{g_0}&=&
\frac{(1-\rho) (1 +\rho)^3 }{2g_0 (1 + (1+\rho)^3 a_2(\lambda))^2}\nonumber\\
&&\times[\rho(1+ (1+\rho)^3 a_2(\lambda))a_1'(\lambda) \nonumber\\
&&- (1 +\rho(1+\rho)^3 a_1(\lambda))a_2'(\lambda)]. \nonumber
\eea

\begin{figure}[htb]
\includegraphics[width=\columnwidth]{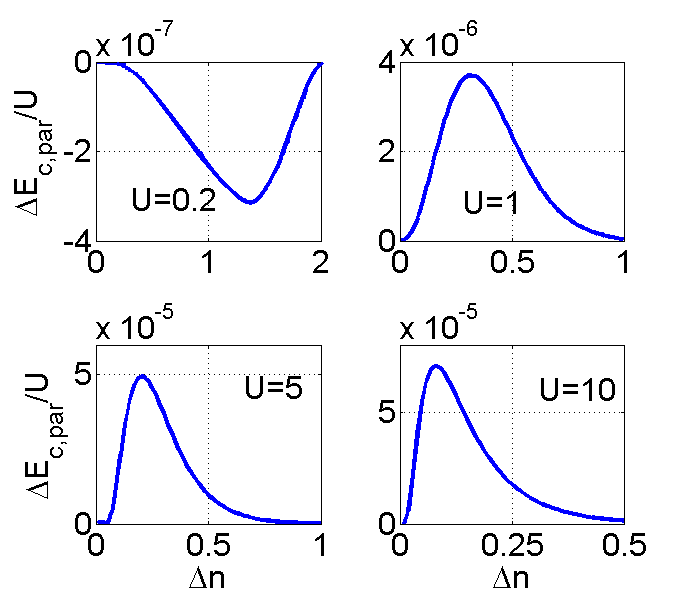}
\caption{Error in $E_{\rm{{\sss C},par}}(\rho)/U$ for different $U$ and $2\,t=1$.}
\label{deltaEcU}
\end{figure}
The new $F_{\rm{par}}$ and $E_{{\sss C},\rm{par}}$ are then obtained by using $g_1$ in Eqs. (\ref{Fung})
and (\ref{Ec1}). Using $g_1$,  $\partial E_{{\sss C},\rm{par}}/\partial g \neq 
0$ still, but the  error with $g_1$ is much lower than with $g_0$.
We plot the the relative error, $(E\c-E_{{\sss C},\rm{par}})/U$ for several $U$ in Fig. \ref{deltaEcU}. The maximum relative error %, $(E\c-E_{\sss {C},\rm{par}})/U$,
is reduced by almost two orders of magnitude (from
$2\times 10^{-3}$ to $5\times 10^{-5}$) in the region $U\approx 2-6$, $\dn\approx 0.25$,
where $g_0$ has the largest error.  The other regions are also improved.
For $(T\c-T_{{\sss C},\rm{par}})/U$ and $(U\c-U_{{\sss C},\rm{par}})/U$ the improvement is just of one order of magnitude (from $2\times 10^{-2}$ to $2\times 10^{-3}$ in both cases relative to the maximum), with different sign, so there is an error cancellation that yields the larger reduction of the $E\c$ error.
We anticipate that $g$ could be improved even further by iteration.

\begin{figure} [htb]
\includegraphics[width=\columnwidth]{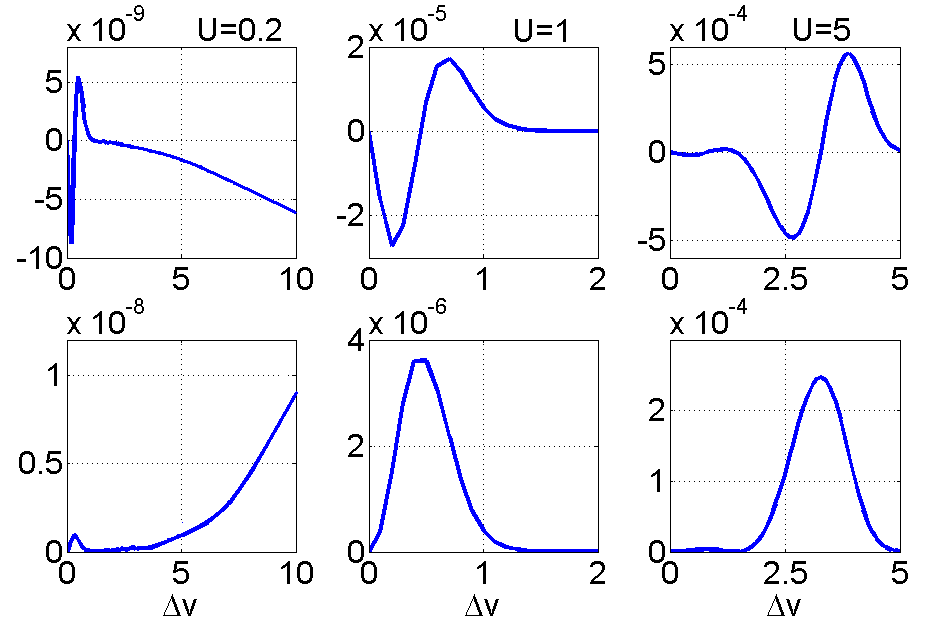}
\caption{
Top row: Error in density as a function of $\deps$.
Bottom row: Error in ground-state energy as a function of $\deps$ and $2\,t =1$. }
\label{Dngspar}
\end{figure}

To test the validity of our parametrization, we use it in the KS scheme to
calculate the correlation energy of our Hubbard dimer {\em self-consistently}.
If our parametrization were perfect, we would recover the exact densities and energies
from our KS calculation without having to solve the many-body problem.
These are plotted in Figs. \ref{Dngspar}, together with the absolute errors committed by the parametric function(al).
Notice that in Figs. \ref{Ec} and \ref{Tc} the results
obtained from the parametric function(al) are indistinguishable from the exact results.
We recommend the use of $g_0$ for routine use, and $g_1$ for improved accuracy.
We hope the methodology developed here might prove useful to improve accuracy
of correlation functionals in other contexts\cite{S15}.

We can define the starting point of our parameterization in a multitude of ways.  In this section
we defined it such that the parameter corresponds to the hopping term.  Another possible
choice favors the electron-electron term.  Define,
\ben
f_2(f,\rho)=-2\,t\,\sqrt{1-f}\,\left(\sqrt{f+\rho}+\sqrt{f-\rho}\right)+U\,f.%\nonumber
\een
Another choice captures the asymmetric limit.  Define,
\ben
f_3(l,\rho)=-2\,t\,\sqrt{2\,l-l^2-\rho^2}+U\,\frac{l^2+\rho^2}{2\,l}.%\nonumber
\een
Then,
\ben
F(\rho) = \min_f f_2(f,\rho) = \min_l f_3(l,\rho)
\een 
These also yield high order polynomial equations when minimized.  The present
parameterization, Eq. (\ref{g0}), is quantitatively superior for nearly all values of $U$, and $\dv$ of interest.

\sec{Approximations}

The usefulness of KS-DFT derives from the use of approximations for the XC functional,
not from the exact XC which is usually as expensive to calculate as direct solution of the
many-body problem (or more so).  While the field of real-space DFT is deluged by
hundreds of different approximations\cite{MOB12}, (relatively few of which are used in routine calculations\cite{PGB15}) 
few approximations exist that apply directly to the Hubbard dimer.
The two we explore here are illustrative of many general principles.

\ssec{Mean-field theory: Broken symmetry}
\label{MF}

Since time immemorial, or at least the 1930's, folks have realized the limitations
of restricted HF solutions for strongly correlated multi-center problems, and 
performed broken-symmetry calculations\cite{CF49}.  For example, in many-body theory, Anderson solved
the Anderson impurity model for a magnetic atom in a metal\cite{A61} by allowing symmetry
breaking, several years before Kondo's ground-breaking work\cite{Kb64}.   In quantum chemistry,
Coulson and Fischer identified the Coulson-Fischer point of the stretched H$_2$ molecule
where the broken symmetry solution has lower energy than the restricted solution\cite{CF49}.
Modern quantum chemists like to spin purify their wavefunctions, but DFT hardliners\cite{PSB95}
claim the broken-symmetry solution is the `correct' one (for an approximate functional).
The exact KS functional, as shown in all previous sections, yields the exact energy and
spin densities, while remaining in a spin singlet.

If we do not impose spin symmetry, the effective potential in mean-field theory becomes (Sec \ref{sec:hubbard}):
\ben
v\eff_{i\sigma}=v_i + U\, \n_{i\bar\sigma},
\een
with $\sigma=+1$ for spin up, $\sigma=-1$ for spin down and
$\bar\sigma=-\sigma$, because the change in the effective field is caused by the
other electron.  
Writing 
$n_i=n_{i,\uparrow}+n_{i,\downarrow}$,
$m_i=n_{i,\uparrow}-n_{i\downarrow}$ and $\Delta m=m_1-m_2$, and
defining 
\ben
\dv\eff_\sigma=\deps-\frac{U}{2}\,(\dn-\sigma\,\Delta m),
\label{veffs}
\een
and
\ben
t\eff_\sigma = t \sqrt{1 + (\dv\eff_\sigma/2\,t)^2},
\label{teffs}
\een
we find the eigenvalues are:
\ben
\label{MFDef1}
e^{MF}_{\pm,\sigma}=\frac{U}{4}\,(N-\sigma\,M)\pm\,\frac{t\eff_{\bar\sigma}}{2},
\een
where $N=2$ is the number of particles and $M$ is the total magnetization.
We find the ferromagnetic solution ($M=2$) to be everywhere above the 
antiferromagnetic solution ($M=0$), and for $M=0$:
\ben
E=\frac{U}{2}\,(1-\frac{\dn^2-\Delta m^2}{4})-\half (t\eff\up+t\eff_\downarrow),
\een
where $\dm=0$ is the paramagnetic (spin singlet) solution, and corresponds
to our original mean-field or restricted Hartree-Fock solution.
We minimize this energy with respect to $\dn$ and $\Delta m$, given by
\ben
\dn=\sum_\sigma \frac{\dv\eff_\sigma}{t\eff_\sigma},~~~
\dm=\sum_\sigma \sigma \frac{\dv\eff_\sigma}{t\eff_\sigma},
\label{scf}
\een
These antiferromagnetic (AFM) self-consistency equations always have the trivial solution $\dm=0$, which corresponds 
to the restricted MF solution(RHF). However, there exists a
non-trivial solution $\dm\ne 0$ for sufficiently large  values of $U$. 

\begin{figure}[htb]
\includegraphics[width=.9\columnwidth]{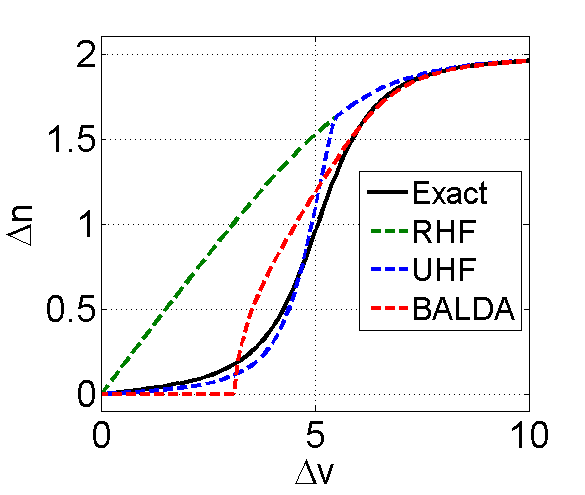}
\caption{
Plots of $\dn$ for HF and BALDA as a function of $\dv$ for $U=5$ and $2\,t=1$.
The crossover from the charge-transfer to the Mott-Hubbard regime happens at $U\approx\dv$.
}
\label{dndmU}
\end{figure}

In Fig. \ref{dndmU}, we plot $\dn$ and $\dm$ for both restricted and unrestricted
HF solutions for $U=5$.  The solutions coincide for large $\dv$, but
below a critical value of $\dv$, they differ.  The UHF solution has
a significantly lower $\dn$, which is much closer to the exact $\dn$.

\begin{figure}[htb]
\includegraphics[width=.9\columnwidth]{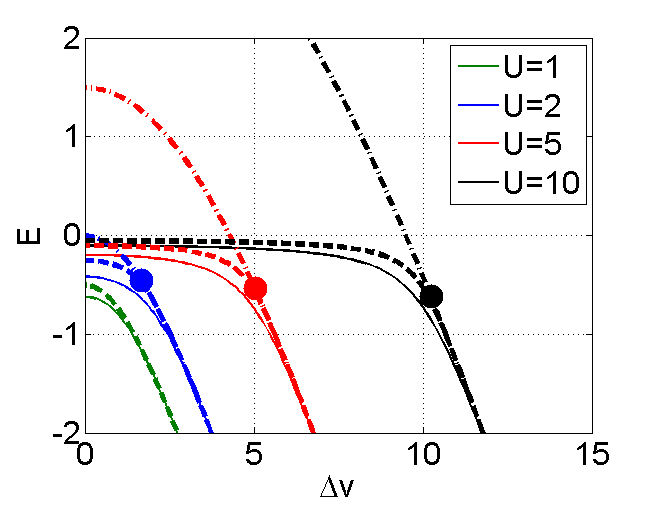}
\caption{Ground-state energy of the unrestricted Hartree-Fock (thick dashed line),
restricted Hartree-Fock (dot dashed line), and
exact ground-state (thin solid line) of the Hubbard dimer as a function of $\dv$ for
several values of $U$ and $2\,t=1$. The dot shows the Coulson-Fischer point at which
the symmetry breaks spontaneously.
For smaller $\dv$ the UHF energy is below RHF while for larger $\dv$ they are the same.}
\label{EUHF}
\end{figure}

In Fig. \ref{EUHF}, we plot the energies, showing that the UHF solution
does not rise above zero, and mimics the exact solution rather closely.
For large $U$, at $n_1=1$, we can compare results analytically:
\ben
E\to
\frac{U}{2} - 2\,t~~(\rm{RHF}),
~-\frac{2\,t^2}{U}~~(\rm{UHF}),
~-\frac{4\,t^2}{U}~~(\rm{exact})
\een
confirming that the UHF energy is far more accurate than the RHF energy, and
recovers the dominant term in the strongly correlated limit.
Note that the symmetric case is atypical:  The constant terms vanish,
both exactly and in UHF, so the leading terms is $O(1/U)$, and its coefficient
in UHF is underestimated by a factor of 2.
The slope of the exact result is two times larger than UHF.
Of course, the exact solution is a spin-singlet, so the symmetry of
the UHF solution is incorrect, but its energy is far better than that of RHF.
This is called the symmetry dilemma in DFT\cite{PSB95}: Should I impose
the right symmetry at the cost of a poor energy?  Note that the exact KS wavefunction
is also a singlet, so a broken-symmetry DFT solution produces the wrong
symmetry for the KS wavefunction.

\ssec{BALDA}
\label{BALDA}

In real-space DFT, the local density approximation (LDA) was first suggested by
Kohn and Sham\cite{KS65}, in which the XC energy is approximated at each point
in a system by that of a uniform gas with the density at that point.
Another way to think of this is that one decides to make a local approximation,
and then chooses the uniform gas XC energy density to ensure exactness in the
uniform limit.
On the lattice, we must switch our reference system to
incorporate Luttinger-liquid correlations instead of Fermi-liquid correlations\cite{Hb81}.
The infinite homogeneous Hubbard chain plays the role of the uniform
gas.   This can be solved exactly via Bethe ansatz\cite{LW68}, and
the corresponding LDA was first constructed and tested in Ref. \cite{SGN95}.
Later, 
Capelle and collaborators\cite{LOC02,CLSO02,LSOC03,XPTC06,FVC12} used the exact Bethe ansatz solution 
to create an explicit parameterization for the energy per site, and called this
Bethe Ansatz LDA, or BALDA.

Since its inception, BALDA has been applied to many different problems including disorder
and critical behavior in optical lattices\cite{X08,CQH09}, spin-charge separation\cite{V12,V14}
and effects of spatial inhomogeneity\cite{SLMC05,LMC07} in strongly correlated systems, 
confined fermions both with attractive and repulsive interactions\cite{CC05}, current DFT on
a lattice\cite{AS12}, electric fields and strong correlation\cite{AS10}, and various critical
phenomena in 1-D systems\cite{ABXP07,FHB12}.
Extensions to include spin-dependence (BALSDA) have been
principally used for studying density oscillations\cite{X12,VFCC08}, and fermions in 
confinement\cite{XA08,X13,HWXO10}.  A thermal DFT approximation on the lattice has been 
constructed using BALDA\cite{XCTK12}.
BALDA has also been used as an adiabatic approximation in TD-DFT to calculate excitations\cite{V08,LXKP08,KS11,UKSS11,VKPA11,KUSS12}
and also transport properties\cite{KSKV10,VKKV13}, as well as using BALDA as a gateway to 
calculate time-dependent effects in 3-D\cite{KPV11}.
There has been significant interest in using BALDA to understand the derivative discontinuity 
in both DFT and TD-DFT\cite{XPTC06,KSKV10,XCTK12,YBL14}.
Additionally, the BALDA approach has been developed for other BA-solvable fermionic lattice systems aside from the 
Hubbard model\cite{XPAT06,APXT07,SDSE08,MTb11}, such as the Anderson model\cite{BLBS12,LBBS12,KS13}, as well as bosonic systems\cite{HCb09,WHZ12,WZ13}.

\begin{figure}[htb] 
\includegraphics[width=\columnwidth]{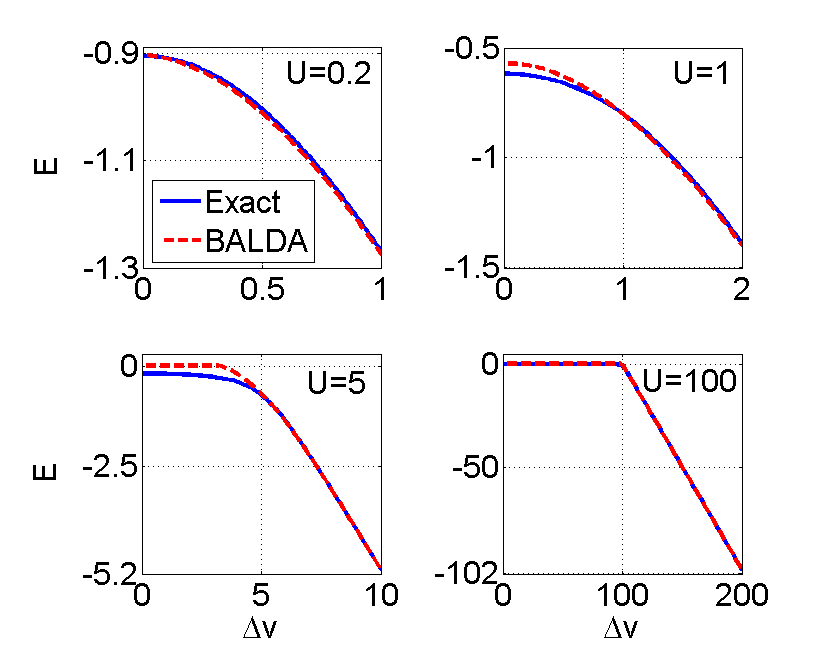}
\caption{Ground-state energy versus $\deps$ for several $U$, 
with $2\,t=1$. 
The BALDA energies are evaluated self-consistently.}
\label{E0BAdv}
\end{figure}
We use here the semi-analytical approach to BALDA\cite{LSOC03, XPTC06} where the expressions are
given in Appendix \ref{BALDAder}. 
In Fig. \ref{E0BAdv} we plot the BALDA ground-state 
energy as a function of $\dv$ for several values of $U$.
At first glance, it seems to do a good job in all regimes.  
In particular, for either very weak correlation ($U=0.2$) or very strong
correlation ($U=100$), it is indistinguishable from the exact curves.
However, for moderate correlation ($ 1 \lesssim U \lesssim 5$) where $\dv \lesssim U$, it appears
to significantly underestimate the magnitude of $E$.

Even for the strong correlation regime, its behavior is not quite correct.  For the
symmetric case:
\ben
E^{BA}\simeq 2\,t\left(\frac{4}{\pi}-1\right) > 0 ~~~~~~~ (U \gg 2\,t)
\een
Thus, for $\deps=0$ and $U=100$ in Fig. \ref{E0BAdv}, BALDA is in serious error, but this
cannot be seen on the scale of the figure.  The origin of this error is easy to understand.
BALDA's reference system is an infinite homogeneous chain, and we are applying it to a finite inhomogeneous dimer.  
The error is in the correlation kinetic energy, which comes from the difference
between the exact and KS kinetic energies.  The tight-binding energy for an infinite
homogeneous chain is different from that of the dimer, and this difference is showing
up (incorrectly) in the correlation energy.

\ssec{BALDA versus HF}
\label{BALDAvsHF}

\begin{figure}[htb]
\includegraphics[width=\columnwidth]{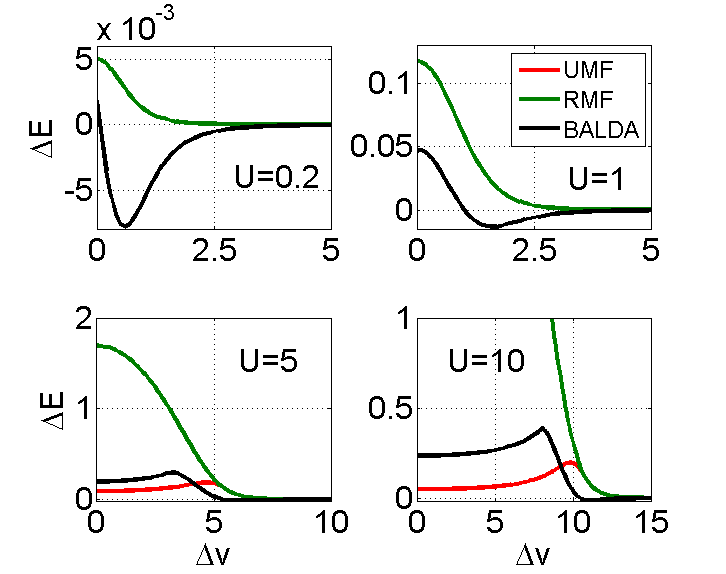}
\caption{
Plots of the RMF, UMF, and BALDA $\Delta E = E^{\rm{approx}} - E^{\rm{exact}}$ as a function
of $\dv$ for $U = 0.2,\,1,\,5,$ and $10$.  For small $U$ the RMF and UMF results are
indistinguishable.
Here $2\,t=1$.}
\label{Eapp}
\end{figure}
Lastly we compare BALDA and both the restricted and unrestricted Hartree-Fock approximations.
In Fig. \ref{Eapp}, we plot the errors made in the ground-state energy of all three
approximations.  For $U \leq 1$, HF does not break symmetry, and so UHF=RHF.  For
very small $U$, the energy error is comparable to HF.  For $U=1$, BALDA is better than
HF.  For larger $U$, UHF produces a lower energy than HF, and almost everywhere is 
more accurate than BALDA.  The sole exception is at precisely $U\approx\deps$, where
BALDA is much better.   
In Fig. \ref{dndmU}, we compare BALDA and UHF densities to the
exact density for $U=5$ as a function of $\dv$.  Although BALDA does not
have a symmetry-breaking point, it unfortunately has a critical value
of $\dv$ where $\dn$ vanishes incorrectly.  This is the origin of the
cusp-like features in the BALDA energies of Figs. \ref{E0BAdv} and \ref{Eapp}.   In fact,
the BALDA density appears somewhat worse than UHF for most $\dv$.
But keep in mind that the main purpose of BALDA is to produce
accurate energies {\em without}
the artificial spin-symmetry breaking of UHF.

\sec{Fractional particle number}
\label{frac}

We will now show a way that one {\em can} extract the physical gap from 
ground-state DFT.  This is done simply by changing the number of
electrons, but now continuously, rather than just at integers.
In fact, we already used this technology implicitly in Sec \ref{gap}, but
here we make this much more explicit.

\ssec{Derivative discontinuity}

An extremely important concept in DFT is that of the derivative discontinuity
\cite{PPLB82,PL83,SG87,MCY08,CMYb08,MCY09,KSKV10,YCM12,MC14}.
This is most famous for its implication for the Kohn-Sham gap of a solid, ensuring
that the gap (in general) does {\em not} match the true fundamental (or charge) gap of
the solid, as we saw in Sec. \ref{gap}.
The expression
itself refers to a plot of ground-state energy versus particle number
$N$ at zero temperature.   In seminal work\cite{PPLB82,PL83,Pb85}, it was shown that
$E(\cN)$ consists of straight-line segments between integer
values, where $\cN$ is a real variable, where all quantities are now expectation
values in a grand-canonical ensemble at zero temperature:
\ben
E(\cN) = (1-w)\, E(N) + w\, E(N+1),
\een
and
\ben
\n_\cN (\br)= (1-w)\, \n_N(\br) + w\, \n_{N+1}(\br),
\een
where $\cN= N + w$, i.e., both energy and ground-state density are piecewise
linear, with a sudden change at integer values.  

Then the chemical potential is
\bea
\mu = dE/d\cN &=& - I~~~~~~~~ (\cN < N)\nonumber\\
&=& - A~~~~~~~~ (\cN > N).
\eea
When we evaluated everything at $N=2$ in Sec. \ref{gap}, we really meant $N=2^-$.
Then Janak's theorem\cite{J78} shows that, for the KS system,
\bea
\mu = dE/d\cN &=& \ehomo ~~~~~~~~ (\cN < N)\nonumber\\
&=& \elumo ~~~~~~~~ (\cN > N)
\eea
This is the proof of the equivalence of $I$ and $-\ehomo$.  

Because the energy is in straight-line segments, the
slope of $E(\cN)$, the chemical potential, $\mu(\cN)$, jumps discontinuously
at integer values.   Hence the name, derivative discontinuity.  The jump in $\mu$
across an integer $N$ is then $E_g=I-A$, the fundamental gap.  In the KS system, 
since the energy is given in terms of orbitals and their occupations, that
jump is simply the KS HOMO-LUMO gap, $E_{gs}$.  
Since the KS electrons have the non-interacting
kinetic energy, and the external and Hartree potentials are continuous functionals
of the density, the difference is an XC effect.  Moreover, it implies that
$v\xc$ jumps by this amount 
as one passes through $N$, an integer.

For solids, addition or removal of a single electron has an infinitesimal
effect on the density, but the XC discontinuity shifts the conduction band upward
by $\Delta\xc$ when an electron is added, contributing to the true gap.   Since local and semilocal
approximations to XC are usually smooth functionals of the density, they produce no
such shift.   They {\em do} yield accurate approximations to the KS gap of a solid,
but {\em not} to the gap calculated by adding and removing an electron, because of
this missing shift.   Thus we have no general procedure for extracting accurate gaps
using LDA and GGA.  An important quality factor in more sophisticated approximations is
whether or not they have a discontinuity.  Orbital-dependent functionals, such as 
exact exchange (EXX in OEP)\cite{SH53,TS76,KLI92,G05,YW02,KK08} or self-interaction corrected LDA (SIC)\cite{PZ81,HHL83,JGGP94,PASS07,PRP14}, often capture
effects due to the discontinuity quite accurately.

\ssec{Hubbard dimer near integer particle numbers}

\begin{figure}[htb] 
\includegraphics[width=0.9\columnwidth]{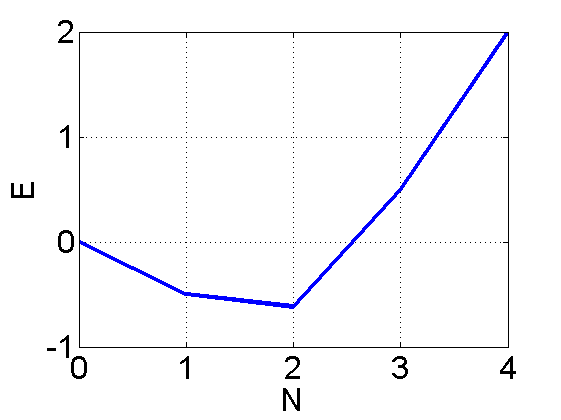}
\caption{
Plot of $E(\cN)$ for $U=1$, $\deps=0$ and $2\,t=1$.
}
\label{EvsN}
\end{figure}
In Fig. \ref{EvsN}, we plot $E(\cN)$ for our Hubbard dimer.  Real-space
curves have always been found to be convex, although this has never
been proven to be generally true.  
The vital part for us is that this equivalence of
the HOMO level and $-I$ links the overall position of the KS
levels to those of the many-body system.  For fixed particle number, only the
KS on-site energy difference is determined by the need to reproduce the exact
site occupancies.  But this condition also fixes the mean value of the KS on-site
energy, $\bar v\s$, which in general is non-zero, even though we chose the actual mean
on-site energy to be zero always.  
In Fig. \ref{cartB}, this is visible in the mean position of the two KS on-site potentials.

\begin{figure}[htb] 
\includegraphics[width=0.9\columnwidth]{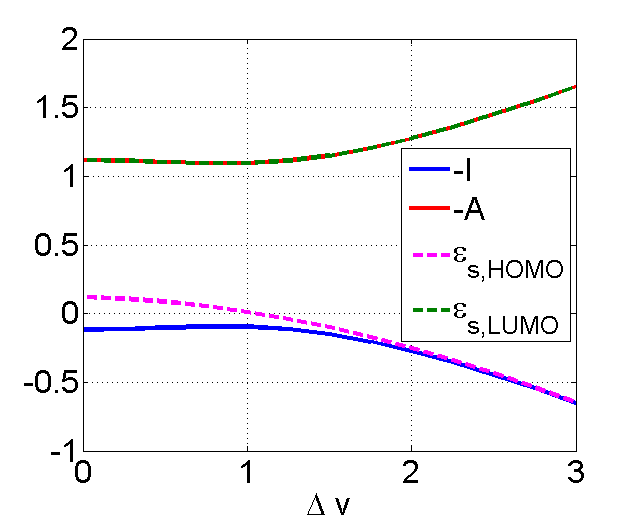}
\caption{
Same as Fig. \ref{N2minusU1} except with $N=2^+$ instead of $N=2^-$.
}
\label{N2plusU1}
\end{figure}
Another way to think about this is that function(al)
derivatives at fixed $\cN$ leave an undetermined constant in the potential, whereas
that constant is determined if the particle number is allowed to change.
We can write many equivalent formulas for the discontinuity[]:
\bea
\Delta\xc & = & \frac{\partial E\xc}{\partial N}\Big|_{N^+}
-\frac{\partial E\xc}{\partial N}\Big|_{N^-},\nonumber\\
&=& \bar v\xc(N^+) - \bar v\xc(N^-),\nonumber\\
&=& \bar v\s(N^+) - \bar v\s(N^-),\nonumber\\
&=& \epsilon\s(N^+) - \epsilon\s(N^-),
\eea
all of which are true.  Thus another way to find the gap from a KS system is to occupy
it with an extra infinitesimal of an electron, and note the jump in potentials or
eigenvalues.  To illustrate this, in Fig. \ref{N2plusU1} we replot Fig. \ref{N2minusU1}, but now for
$N=2^+$, showing that now the LUMO matches $-A$, and the difference between the
HOMO and $-I$ is $\Delta\xc$. 

\begin{figure}[htb] 
\includegraphics[width=0.90\columnwidth]{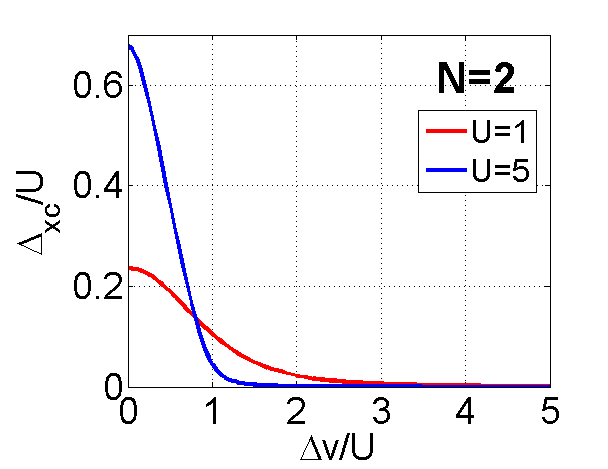}
\caption{Derivative discontinuity as a function of $\dv$ for 
$U=1$, and $U=5$.} 
\label{dxc}
\end{figure}
In Fig. \ref{dxc} we plot $\Delta\xc$ for $N=2$
for various $U$, as a function of $\dv$, scaling each variable by
$U$.  We see that the discontinuity always decreases with increasing
$\dv$.  In fact, the larger $U$ is, the more abruptly it vanishes
(on a scale of $U$) when $\dv > U$.  In this sense, the greater
the asymmetry, the less discontinuous the energy derivative is,
and the KS gap will be closer to the true gap.

\begin{figure}[htb] 
\includegraphics[width=0.90\columnwidth]{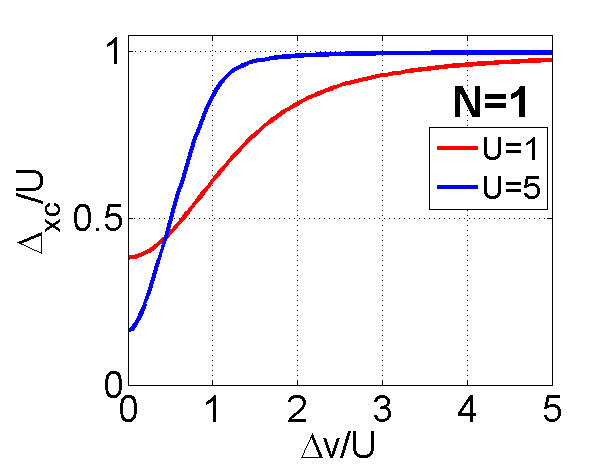}
\caption{Derivative discontinuity for $N=1$ as a function of $\dv$ for 
$U=1$, and $U=5$.} 
\label{dxcN1}
\end{figure}
The situation is reversed when $N=1$, as shown in Fig. \ref{dxcN1}.  Now the
discontinuity grows with increasing $\dv$.  In this case, a large
asymmetry puts the electron mostly on one site.  When an infinitesimal
of an electron is added, it goes to the same site, but paying an energy
cost of $U$.  On the other hand, if $\dv$ is small, the first electron
is spread over both sites, and so is the added infinitesimal, reducing
the energy cost by a factor of 2.  So $\Delta\xc \to U/2$ in the
weakly correlated near-symmetric limit.

\ssec{Discontinuity around $n_1=1$ for $N=2$}

The derivative discontinuity manifests itself in many different aspects of
DFT.  We have already seen how it affects both energies and potentials
as $N$ is continuously moved across an integer.
Here we explore how it appears even at fixed particle number, as correlations
become strong.

For our Hubbard dimer, with any finite $\dv$, if $U \gg \dv$, we know each
$\n_i$ is close to 1.  The overwhelmingly large $U$ localizes each electron
on opposite sites.  In the limit as $U\to\infty$, all fluctuations are
suppressed, and the dimer becomes two separate systems of one electron
each.  For large but finite $U$, and finite $\dv$, one is on the integer
deficient side, and the other has slightly more than one electron.  All the
statements made above about $\cN$ passing through 2
now apply as $n_1$ passes through 1.  

We can see the effects in many of our earlier figures.  In Fig. \ref{ffunctional}, the
slope of $F$ for $U=10$ appears discontinuous at $n_1=1$.  $F$ contains
the discontinuity in both $T\s$ and $E\xc$ in the limit $U\to\infty$.  However,
in reality, this curve is not really discontinuous.  Zooming in on $F$
near $\n_1=1$, one sees that on a scale of $O(1/U)$, $F$ is rounded.

\begin{figure}[htb]
\includegraphics[width=.9\columnwidth]{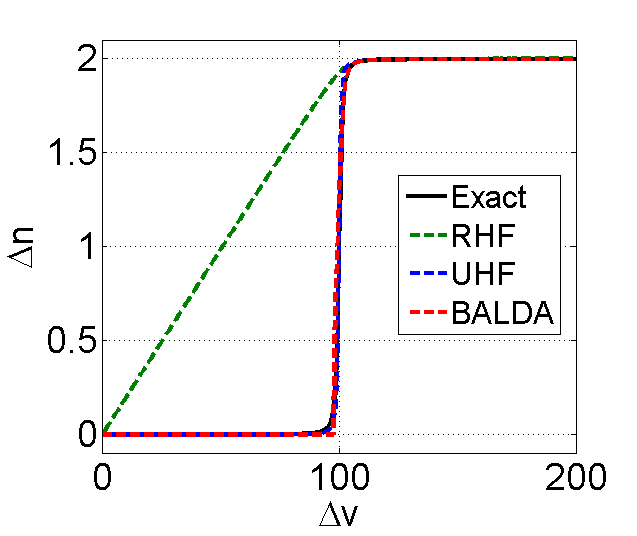}
\caption{Plots of $\dn$ and $\Delta m$ in HF as a function of $\dv$
for $U=100$ ($2\,t=1$), and the BALDA charge density.
The crossover from the charge-transfer to the Mott-Hubbard regime
happens at about $U\approx \deps$.}
\label{dndmU100}
\end{figure}
The classic manifestation already appears in Fig. \ref{dndv}, the occupation
difference as a function of $\dv$.    To emphasize the point, 
in Fig. \ref{dndmU100}, we plot several curves for $U=100$.  
This is the discontinuous change from having
1 particle on each site to 2 on one site that occurs.  
This is important because the common approximate density functionals
miss this discontinuity effect.  Explicit continuous functionals of the density  
cannot behave this way.  For the SOFT case, this is embodied in the HF
curves of Fig. \ref{dvc}:  No matter how strong the value of $U$, these curves
are linear.  In RHF, $\dn$ versus $\dv$ never evolves the sudden step
discussed above, as shown in Fig. \ref{dndmU}.
On the other hand, the BALDA approximation contains
an explicit discontinuity at $\n_1=1$ in its formulas, and so captures
this effect, at least to leading-order in $U$.  
In this sense, both BALDA and UHF capture
the most important effect of strong correlation. 
On the other hand,
as discussed in Sec \ref{BALDAvsHF}, UHF `cheats', while BALDA retains the correct spin
singlet.  If BALDA's effects could be (legally) built into real-space
approximations, they would be able to accurately dissociate molecules,
overcoming perhaps approximate DFT's greatest practical failure.

\begin{figure}[htb]
\includegraphics[width=.9\columnwidth]{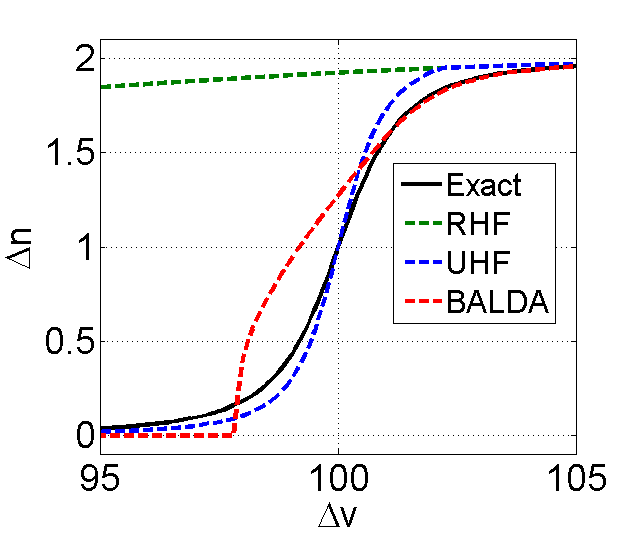}
\caption{Plots of $\dn$ and $\Delta m$ in HF as a function of $\dv$
for $U=100$ ($2\,t=1$), and the BALDA charge density.
The crossover from the charge-transfer to the Mott-Hubbard regime
happens at about $U\approx \deps$.}
\label{dndmU100zoom}
\end{figure}
However, in Fig. \ref{dndmU100zoom}, we simply zoom in on the
region of the plot near $\dv=U$.  In fact, the exact curve is
$S$-shaped, with a finite curvature on the scale of $t$.
Now we see that, although both UHF and BALDA reproduce the discontinuous
effect, the details are not quite right.  UHF is admirably close in
shape to the accurate curve, but its slope is too great
at $n_1=1$.  BALDA is accurate to leading order in $1/U$, and
captures beautifully the region $\dv$ a little larger than $U$,
but is quite inaccurate below that.  The presence of the gap
in the BALDA potentials leads to the incorrect discontinuous
behavior near $\dv = 98$.  But once again we emphasize that
the important feature is that these approximations do capture
the dominant effect, and that BALDA does so without breaking
symmetry.

\sec{Conclusions and Discussion}
\label{conclusion}

So, what can we learn from this exercise in applying DFT methods to the simplest
strongly correlated system?  Perhaps the most important point is that there is
a large cultural difference between many-body approaches and DFT methodology, and
a considerable barrier to communication.  In Sec. \ref{KSmethod}, we saw that even the definition
of exchange is different in the two communities.   The greatest misunderstandings come
not from using different words for the same thing, but rather from using the same
word for two different things.

We can also see that the limitations of DFT calculations are often misunderstood in
the broader community.  For example, the exact ground-state XC functional has a 
HOMO-LUMO gap that does not, in general, match the fundamental gap.  The KS eigenvalues
are not quasiparticle eigenvalues in general, and are in fact, much closer to optical
excitations\cite{SUG98}.   Even the purpose of a DFT calculation is quite foreign to most solid-state
physics.  The modern art of DFT is aimed at producing extremely accurate (by physics
standards) ground-state energies, and the many properties that can be extracted from those,
rather than the response properties that are probed in most solid-state experiments,
such as photoemission.  (Flipping the coin, most quantum chemists would never describe
DFT energies as extremely accurate, as traditional quantum chemical ab initio methods
are hyper accurate on this scale.)

We also mention many aspects that we have {\em not} covered here.  For example, 
time-dependent DFT is based on a distinct theorem (the Runge-Gross theorem\cite{RG84}),
and provides approximate optical excitations for molecular systems\cite{BWG05}.  
The Mermin theorem\cite{M65} generalizes the HK theorem to thermal ensembles\cite{PPGB14}.
There are many interesting features related to spin polarization and dynamics, but
very little is relevant to the system discussed here.  There are also many non-DFT
approaches, such as $GW$, which could be tested on the asymmetric dimer.

We also take a moment to discuss how SOFT calculations can be related to real-space
DFT.   One can easily add more orbitals to each site and create an extended Hubbard
model.   For the H$_2$ molecule, adding just $p_z$ orbitals and allowing them to
scale yields a very accurate binding curve.  But such an extension (beyond one basis
function per site) is extremely problematic for SOFT\cite{H86}, because it is no longer
clear how to represent the `density'.  With 2 basis functions, should one use just
the diagonal occupations, or include off-diagonal elements?  In fact, neither one
is satisfactory, as neither approaches the real-space density functional in the
infinite basis limit.  An underlying important point of DFT is that it is applied
only to potentials that are diagonal in $\br$, i.e., $v(\br)$, and not diagonal
in an arbitrary basis.  This is a key requirement of the HK theorem, and is the
reason why the one-body density $\n(\br)$ is the corresponding variable on which
to build the theory, and why the local density approximation is the starting
point of all DFT approximations.  

This inability to go from SOFT calculations to real-space DFT calculations should
be regarded as a major caveat for those using SOFT to explore DFT.  Here we have
shown many similarities in the behavior of SOFT functionals compared to real-space
functionals.  We have also proven some of the same basic theorems as those used
in real-space DFT.  But any results (especially unusual ones) that are found in
SOFT calculations might not generalize to real-space DFT.   The only way to be sure
is to find a proof or calculation in real-space.  On the other hand, SOFT calculations
can be safely used to illustrate the basic physics behind real-space results\cite{SWWB12}.

Another limitation of SOFT can be seen already in our asymmetric Hubbard dimer.
In a real heterogeneous diatomic molecule, say LiH with a pseudopotential for the
core Li electrons, the values of $U$ would be different on the two sites.  But the
basic DFT machinery only applies if the interaction is the same among all particles.
And even if it applies when both $U$ and $t$ become site-dependent, i.e., a one-to-one
correspondence can be proven, it is unlikely that such studies would yield behavior
that is even qualitatively similar to real-space DFT.

Finally, we wish to emphasize once again the importance of testing ideas on the
asymmetric Hubbard dimer.   Much (but not all) of the SOFT literature tests
ideas on homogeneous cases.  The essence of DFT is the creation of a universal
functional. i.e., $F[\n]$ is the same no matter which specific problem you 
are trying to solve.  The symmetric case is very special in several ways,
and there are no difficulties in applying any method to the asymmetric case.
We hope that some of the results presented here will make that easier.

\begin{acknowledgments}

We thank Fr\'ed\'erik Mila for his kind hospitality at EPFL where this collaboration began.
We also thank our colleagues A. Cohen, P. Mori-S\'anchez and J. J. Palacios  
for discussions on matters related to this article.
Work at Universidad de Oviedo was supported by the Spanish MINECO project FIS2012-34858, and 
the EU ITN network MOLESCO.
Work at UC Irvine was supported by the U.S Department of Energy (DOE), Office of Science,
Basic Energy Sciences (BES) under award \# DE-FG02-08ER46496. J.C.S. acknowledges support
through the NSF Graduate Research fellowship program under award \# DGE-1321846.

\end{acknowledgments}

\appendix

\sec{Exact solution, components, and limits}
\label{energycomp}
In all appendices, we use dimensionless variables for brevity.  Hence $\epsilon = E/2\,t$, 
$u = U/2\,t$, and $\nu = \dv/2\,t$.
Then, the energy of the singlet-ground-state is
\ben
\label{eq:gsenergy}
\epsilon=\frac{2}{3}\left(u-w\sin\,(\theta+\frac{\pi}{6}) \right)
\een
where
\ben
w=\sqrt{3\,[1+\nu^2]+u^2},
\een
and
\ben
\cos(3\theta)=(9(\nu^2 - 1/2) + u^2)u/{w^3}.
\een

The coefficients of the minimizing wavefunction, Eq. (\ref{wf}), are
\bea
\alpha&=&c\,\left(1-\frac{u}{\epsilon}\right),\,\,\,
\beta_{1,2}=c\,\left(u-\epsilon\pm\nu\right),\\
c^{-2} &=&2\left(\nu^2+(\epsilon-u)^2\left(1+\epsilon^{-2} \right)\right).
\eea

The ground-state expectation values of the density difference and of the different pieces 
of the Hamiltonian are
\bea
\label{Dn}
\dn&=&4\,c^2\,\nu\,(u-\epsilon)\\
\label{von}
V&=&-\deps\,\dn/2,\\
\label{hop}
T&=&4\,c^2(\epsilon-u)^2/\epsilon,\\
\label{coul}
V\ee&=&4\,c^2\,t\, u\,\left((\epsilon-u)^2+\nu^2\right).
\eea

For fixed asymmetry $\nu$, we can expand $\epsilon$ in the weakly and strongly correlated limits:
\ben
\epsilon^w = -\sqrt{1+ \nu^2} \left(1- (\frac{1}{2} + \nu^2)\tilde{u} + (\frac{1}{4} + \nu^2)\frac{\tilde{u}^2}{2} + 
\nu^4 \frac{\tilde{u}^3}{2} \right)
\een
where $\tilde{u} = u/(1 + \nu^2)^{3/2}$. In the strongly correlated limit:
\ben
\epsilon^{st} = -u^{-1} +(1 - \nu^2) u^{-3} + O(u^{-5}).
\een
We can also expand for fixed $u$ around the symmetric limit:
\ben
\epsilon^{sym} = \frac{1}{2} (u - r)  + \frac{u-r}{r(u+r)} \nu^2,
\een
where $r=\sqrt{u^2 + 4}$. And the asymmetric limit:
\ben
\epsilon^{asy} = -\nu + u - (2\nu)^{-1} - u/2 \nu^{-2} + (1-4 u^2) (2\nu)^{-3}.
\een

\sec{Many limits of $F(\Delta n)$}
\label{limits}

In this appendix we derive the limits that our parameterization in Section \ref{param} satisfies.
Minimizing $\tilde{F}$ of Eq. (\ref{Fung}) with respect to $g$, we obtain a sextic equation for $g$:
\bea
\label{polynomialg}
(4+u^2)\,g^6/4+(\rho^2\,(3+u^2)-1)\,g^4+\nonumber\\
2\,u\,\rho^2\,g^3+\rho^2\,(\rho^2\,(3+u^2) -(2+u^2))\,g^2-\nonumber\\
2\,u\,\rho^2\,(1-\rho^2)\,g-\rho^4\,(1-\rho^2)=0
\eea
where we define $\rho = \dn/2$.
The solution defines $g_m(\rho)$, and $F(\rho) = F(g_m(\rho),\rho)$.  Next we expand in several
limits.
and $F[U,\rho] = \tilde{F}[U,\rho,g_m]$.
However, equation (\ref{polynomialg}) can not be solved analytically in general.

\ssec{Expansions for $g(\rho,u)$}
We expand $g$ in 4 different limits, which are built into $g_0$ of Eq. (\ref{g0}) in Section \ref{param}.

The weakly correlated limit corresponds to $u\ll 1$. We thus expand $g(\rho,u)$ in powers of $u$
for fixed $\rho$,
\ben
g(\rho,u)=\sum_{n=0}^\infty\,g^{(n)}(\rho)\, u^n/n!,
\een
and insert the expansion into Eq. (\ref{polynomialg}). The coefficients $g^{(n)}$ are found by canceling each term order by order in Eq. (\ref{polynomialg}), yielding
\bea
\label{weakg}
g^{(0)}&=&\sqrt{1-\rho^2},\,\,\,\,\,\,\,\, g^{(1)}=0,\\
g^{(2)}&=&-\frac{(1-\rho^2)^{5/2}}{4}, \,\,\,\,\, g^{(3)}=\frac{3}{4}\,\rho^2\,(1-\rho^2)^3,\nonumber\\
g^{(4)}&=&\frac{9}{16}\,(1-\rho^2)^{7/2}\,(1+7\,\rho^2-24\,\rho^4).\nonumber
\eea
Notice that $n_{1,2}=1\pm\rho$ so that to first order in $U$, Eq. (\ref{Fung}) yields the non-interacting kinetic energy functional of Eq. (\ref{tsdimer}).

For strongly correlated systems, we expand $g$ in powers of $1/u$ while holding $\rho$ fixed 
\ben
g(\rho,u)=\sum_{n=0}^\infty\,\tilde{g}^{(n)}(\rho)\, u^{-n}/n!,
\een
and substitute back into Eq. (\ref{polynomialg}) to find the coefficients. The result is
\bea
\label{strongg}
\tilde{g}^{(0)}&=&\sqrt{2\,\rho\,(1-\rho)},\,\,\,\,\,
\tilde{g}^{(1)}=\frac{1-\rho}{2},\\
\tilde{g}^{(2)}&=&-\frac{3\,(3\,\rho-1)}{8\rho}\,\tilde{g}^{(0)}.\nonumber
\eea
Notice that this expansion breaks down at the symmetric point $\rho=0$.

The other kind of limit keeps $u$ fixed.  The symmetric
limit is equivalent to $\rho\to0$. We expand $g$ in powers of $\rho$ while holding $u$ fixed.
\ben
g(\rho,u)=\sum_{n=0}^\infty\,\bar{g}^{(n)}(u)\, \rho^n/n!,
\een
and substitute back into Eq. (\ref{polynomialg})  to find the coefficients. The result is
\ben
\bar{g}^{(0)}=r^{-1},\,\,\,\,\,\bar{g}^{(2)}=\frac{1}{2}\left(u^2+\frac{u^2/2\,(u^2/2+1)-1}{r}\right)
\een
where $r=\sqrt{1+(u/2)^2}$.

The asymmetric limit is equivalent to $\rho\to1$. We expand $g$ in powers of $\bar{\rho} = 1-\rho$ for fixed $u$:
\ben
g(\rho,u)=\sum_{n=0}^\infty\,\tilde{\bar{g}}^{(n)}(u)\, \bar{\rho}^n/n!,
\een
and substitute back into Eq. (\ref{polynomialg}). The result is
\bea
\tilde{\bar{g}}^{(1/2)}&=&\sqrt{\pi/2},\,\,\,\,\,\,\,
\tilde{\bar{g}}^{(3/2)}=-3 \tilde{\bar{g}}^{(1/2)}/8\\
\tilde{\bar{g}}^{(5/2)}&=&\left(\frac{1}{16}+u^2\right)\,5 \tilde{\bar{g}}^{(3/2)} \nonumber\\
\tilde{\bar{g}}^{(3)}&=&12\,u^3.\nonumber
\eea

\def\weak{^{w}}
\def\str{^{str}}
\def\sym{^{sym}}
\def\anti{^{asym}}
\ssec{Limits of the correlation energy functional}
\label{correxp}

Now that we have expressions for $g$ in all four limits we can use our expression for $F$, eq. 
(\ref{Fung}), $T\s$, and $U\H$ to compute $E\c$ in each regime:
\ben
e\c=-g+u h(g,\rho)-\frac{u}{2}\,(1+\rho^2)+\sqrt{1-\rho^2}.\nonumber
\een
where $h(g,\rho)$ is defined in Eq. (\ref{h}). Then, as $u\to0$, 
$e\c\to e\c\weak$, where
\ben
\label{Ecweak}
e\c\weak(\rho) =-\frac{u^2}{8}\,(1-\rho^2)^{5/2}\left(1-u\,\rho^2\,\sqrt{1-\rho^2}\right).
\een
Similarly, as $u\to\infty$, $e\c\to e\c\str$, where
\ben
\label{Ecstrong}
e\c\str(\rho) = -\frac{u}{2}\,(1-\rho)^2
+\sqrt{1-\rho}\left(\sqrt{1+\rho}-\sqrt{2\,\rho}\right)-\frac{1-\rho}{4\,u}.
\een
An alternative expansion is to fix $u$ and expand in $\rho$.
As $\rho\to0$, $e\c\to e\c\sym$, where
\bea
\label{Ecsym}
e\c\sym(\rho) &=&
1-\sqrt{1+\left(\frac{u}{2}\right)^2}\\
&+&\rho^2\left(\!\left(\frac{u}{2}\right)^3\!-\frac{1}{2}+ \sqrt{1+\left(\frac{u}{2}\right)^2} \!\left(\frac{1}{2}+\left(\frac{u}{2}\right)^2\right)\!\right).\nonumber
\eea
As $\rho \to 1$, $e\c\to e\c\anti$, where
\ben
\label{Ecasymm}
e\c\anti(\rho)=u^2\,\bar{\rho}^{5/2}\,\left(-\frac{1}{\sqrt{2}}+u\,\sqrt{\bar{\rho}}\right).
\een
where $\bar{\rho} = 1-\rho$.

\ssec{Order of limits}
\label{doublim}

Finally, we look at how these expressions behave when both parameters are extreme.
The weakly correlated limit has no difficulties near the symmetric point:
\bea
e\c\weak(\rho\to 0)&=&e\c\sym(u\to 0)\nonumber\\
&=&-\frac{u^2}{8}\,\left(1-\frac{5\,\rho^2}{2}\right)+\frac{u^3\,\rho^2}{8}.
\eea
In the asymmetric limit, there are also no problems:
\bea
e\c\weak(\rho\to 1)&=&e\c\anti(u\to 0)\nonumber\\
&=&u^2\,\bar{\rho}^{5/2}\,\left(-\frac{1}{\sqrt{2}}+u\,\sqrt{\bar{\rho}}\right).
\eea
Thus, the expansion in powers of $u$ is well-behaved, and there are no difficulties
using it for sufficiently small $u$.   In the symmetric case, one sees explicitly that
the radius of convergence of the expansion is $u=2$.

On the other hand, the strong coupling limit is more problematic.  Expanding the
strong-couping functional around the symmetric limit, we find
\ben
e\c\str(\rho\to 0) = -\frac{u}{2}+1-\frac{1}{4\,u}-
\sqrt{2\,\rho}+\rho\, 
\left(u+\frac{1}{4\,u}\right),
\een
while reversing the order of limits yields:
\ben
e\c\sym(u\to\infty) = -\frac{u}{2}+1-\frac{1}{u}-
\frac{\rho^2}{2}\,\left(1-u-\frac{1}{2\,u}-
\frac{u^3}{2}\right).
\een
Note the difference beginning in the third terms, i.e., at first-order in $1/u$, even
for $\rho=0$.  Thus for the Hubbard dimer, approximations based on expansions around the
strong-coupling limit are likely to fail for some values of the density.

\sec{Proofs of Energy Relations}
\label{proofs}

Using the notation established in Section \ref{param}, we prove some simple relations
about the energy and its components.  Start with the general expression for the 
energy, Eq. (\ref{Fung}) and (\ref{h}),
\ben
\epsilon = \min_{\rho,g}\left[ -g + u h(g,\rho) - \nu \rho \right].
\een
First take $\rho \to 0$.  The second term reduces to $u \left(1-\sqrt{1-g^2}\right)/2$.  Then let 
$g\to0$, resulting in $h\to0$.  This yields $\epsilon\to0$ and therefore the exact $\epsilon \leq 0$.
This process corresponds to choosing a trial wavefunction, and by Rayleigh-Ritz, the  
ground-state wavefunction will produce a value equal to or below the trial result.

In Hartree-Fock, $g$ reduces to $g_{\rm{HF}} = \sqrt{1-\rho^2}$.  Then, 
\ben
\epsilon\HF= \min_{\rho} \epsilon(g_{\rm{HF}}(\rho),\rho) \geq \epsilon.
\een 
This shows that $\epsilon\c\trad = \epsilon - \epsilon\HF \geq 0$, as in Fig. (\ref{HFgs}).  
The minimization can be performed analytically though it involves solving the quartic polynomial
\ben
\frac{\rho}{\sqrt{1-\rho^2}} + u\,\rho - \nu =0.
\een

Similarly, a DFT exact exchange (EXX) calculation is defined by 
\ben
\epsilon\EXX = \epsilon(g_{\rm{HF}}(\rho_m),\rho_m) \geq \min_{\rho} \epsilon(g_{\rm{HF}}(\rho),\rho)
\een 
where $\rho_m$ is the minimizing density for the many-body problem.  This yields 
$\epsilon\c^{\rm{DFT}} = \epsilon - \epsilon\EXX$, and $\epsilon\c\trad \geq \epsilon\c^{\rm{DFT}}$\cite{GPG96}.

For the kinetic energy alone, $t = -g(\rho_m)$, and
\ben
t\s = \min_{u\to0,\rho}[-g(\rho)] = - \sqrt{1-\rho^2}.
\een
This results in $t\c \geq 0$ since the KS occupation difference 
is defined to minimize the hopping energy.  This combined with the above implies $u\c\leq0$,
as in Eq. (\ref{corrineq}). 

For the adiabatic connection integrand, take a derivative of Eq. (\ref{Uc1}):
\ben
\frac{d u\c\l}{d\lambda} = \frac{u\c(\rho,\lambda)}{\lambda} + \lambda u \frac{\partial h}{\partial g} \frac{\partial g}{\partial\lambda}.
\een
The first term is less than zero by definition but the second needs more unraveling.  
To begin, from Eq. (\ref{Fung}),
\ben
\frac{\partial f}{\partial g} = -1 + u \frac{\partial h}{\partial g},
\een
so, at the solution
\ben
\frac{\partial h}{\partial g} = \frac{1}{u}.
\een

For $\lambda$ near 1, Suppose $g(\lambda) \simeq g(1) + (\lambda -1) g'(1)$, and expand 
$\partial h /\partial g|_{g(\lambda)}$ in $g(\lambda)$ around $g(1)$:
\ben
\left.\frac{\partial h}{\partial g}\right|_{g(\lambda)} = 
\left.\frac{\partial h}{\partial g}\right|_{g(1)} +
(\lambda -1) g'(1) \left. \frac{\partial^2 h}{\partial g^2} \right|_{g(1)}
\een
The first term on the left is  $1/(\lambda u) \approx (2-\lambda)/u$.  After some algebra,
\ben
\left.\frac{\partial g}{\partial \lambda}\right|_{\lambda=1} = - \left(
u \left.\frac{\partial^2 h}{\partial g^2}\right|_{g(1)}\right)^{-1}
\een
Since the hopping term of $f$ is linear in $g$, $\partial^2 f/\partial g^2 = \partial ^2 h/\partial g^2$. The energy is a minimum at $g$ so 
$\partial^2 f/\partial g^2 > 0$, thus $\partial g/\partial\lambda >0$.  Together, this 
results in 
\ben
d U\c\l/d\lambda < 0,
\een
the adiabatic connection integrand is monotonically decreasing as seen in Fig. \ref{ACplot}.

\sec{BALDA Derivation}
\label{BALDAder}

For an infinite homogeneous Hubbard chain of density $n=1+x$, the  energy per site (in units 
of $2\,t$) is given approximately by
\ben
\label{eBALDA}
\tilde \epsilon\unif=u\,x\,\theta(x)+\alpha(x,\beta(U))
\een
where $\theta(x)$ is the Heaviside function and
\ben
\alpha(x,\beta)=-\frac{\beta}{\pi}\,\sin\left(\pi\,(1-|x|)/\beta\right)/\pi.
\een
The function $\beta (u)$ varies
smoothly from 1 at $u=0$ to 2 as $u\to\infty$\cite{LSOC03}, and
satisfies
\ben
\alpha(0,\beta)=-4\int_0^\infty\!d\,\xi\,\frac{J_0(\xi)\,J_1(\xi)}{\xi\,\left[1+\exp(u\,\xi))\right]}
\label{betadef}
\een
This simple result is exact as $u\to 0$,
$u\to\infty$ and at $n=1$, and a good approximation (accurate to within a few
percent) elsewhere\cite{LSOC03} to the exact solution via
Bethe ansatz\cite{LW68}.   In principle, $\beta$ depends on $n$, and this dependence has been
fit in later work\cite{FVC12}.  Here, we use the simpler original version of a function of $u$ only.
In fact, the solution to Eq. (\ref{betadef}) can be accurately fit (error below 1\%)
with a simple rational
function,
\ben
\beta^{fit}(u) = \frac{2 + a u + bu^2}{1+ c u + b u^2}
\een
with coefficients $a= 2c - \pi/4$ and $b = (a-c)/\log{2}$ chosen to recover the small-$u$ behavior to first-order, and
the large $u$ behavior to first order in $1/u$, and $c=1.197963$ is fit to $\beta(u)$. 
This is useful for quick implementation of BALDA.

At $u=0$, the hopping energy per site is just
\ben
\label{Tsunif}
\tilde t\s\unif=
-\sin\left(\pi\,(1-|x|)\right)/\pi,
\een
while the Hartree-exchange energy per site is a simple local function:
\ben
\label{EHXunif}
\tilde u\Hx\unif/= u\, n^2/4.
\een
Thus the correlation energy per site is just
\ben
\tilde \epsilon\c\unif = \tilde \epsilon\unif - \tilde t\s\unif -\tilde u\Hx\unif.
\een
The BALDA approximation is then
\ben
\epsilon\xc\BALDA = \tilde \epsilon\xc\unif(n_1,U)+\tilde \epsilon\xc\unif(n_2,U).
\een
Since the exchange is local, BALDA is exact for that contribution, and
only correlation is approximated.   Since
$n_{1,2}=1\pm\dn/2$, $x=\pm \dn/2$ for sites 1 and 2 respectively.
The BALDA HXC energy is then:
\ben
\epsilon\Hxc\BALDA= -2(\alpha(\dn/2,U)-\alpha(\dn/2,0))+u|\dn|/2,
\een
and was inserted into the KS equations (Sec \ref{KSmethod}) to find the results of Sec \ref{BALDA}.

\sec{Mean-Field Derivation}
\label{app:MF}
The MF hamiltonian for the Hubbard dimer can be written in the number
basis $\left|1\sigma,2\sigma\right\rangle $ as follows
\bea
\hat{H}_\sigma^{MF}&=&\left(\begin{array}{cc}
-\Delta v_{\sigma}^{\rm{eff}} & -t \\
-t &\Delta v_{\sigma}^{\rm{eff}}
\end{array}\right)%\\
\eea
with $\sigma=\pm1$ for spin up and down respectively. 
Setting $M=m_1+m_2$ and $N=n_1+n_2$ as 
the total magnetization and particle number of the system, the eigenvalues are
\bea
\label{MFDef}
e^{MF}_{\pm,\sigma}&=&\frac{U}{4}\,(N-\sigma\,M)\pm\,\frac{t^{\rm{eff}}_{\bar\sigma}}{2},\\
t^{\rm{eff}}_\sigma&=&2\,t\sqrt{(\Delta v^{\rm{eff}}_\sigma/2\,t)^2+1},\nonumber\\
\Delta v^{\rm{eff}}_\sigma&=&\deps-\frac{U}{2}\,(\dn-\sigma\,\Delta m).\nonumber
\eea
The total energy of the system is
\bea
\label{EFM}
E^{FM}&=&e_{-,\uparrow}+e_{+,\uparrow}-U\H\\
E^{AFM}&=&e_{-,\uparrow}+e_{-,\downarrow}-U\H,
\eea
where the Hartree term is written as
\bea
U\H&=&\frac{U}{4}\left(n_{1\,\uparrow}\,n_{1\,\downarrow}+n_{2\,\uparrow}\,n_{2\,\downarrow}\right)\nonumber\\
&=&\frac{U}{8}\left(N^2-M^2+\Delta n^2-\Delta m^2\right).
\eea
Depending on whether $E^{AFM}$ is larger or smaller than $E^{FM}$, the ground-state of the system may be ferromagnetic ($N=2$, $|M|=2$) or antiferromagnetic ($N=2$, $M=0$, $|\Delta m|\ge0$). The paramagnetic
state is a specific case of the AFM state with $\Delta m=0$.
Explicitly, for the ferromagnetic state we have the eigenstate energies and self-consistency equations
\bea
\dn&=&\dm=\dv/\sqrt{4\,t^2+\dv^2}\\
e_{\mp,\uparrow}&=&\mp\sqrt{4\,t^2+\dv^2}/2
\eea

On the other hand, the $M=0$ state ($|\dm|>0$ is AFM, $\dm=0$ is PM) corresponds to the eigenvalues,
\ben
e_{-,\uparrow}=(U-t^{\rm{eff}}_\downarrow)/2,~~~
e_{-,\downarrow}=(U-t^{\rm{eff}}_\uparrow)/2,
\een
and self-consistency equations
\ben
\dn=\sum_\sigma \frac{\dv\eff_\sigma}{t\eff_\sigma},~~~
\dm=\sum_\sigma \sigma \frac{\dv\eff_\sigma}{t\eff_\sigma},
\label{scfapp}
\een
and the expressions for $\dv^{\rm{eff}}_\sigma$ and $t^{\rm{eff}}_\sigma$ are given in  Eq. (\ref{MFDef}). 
The self-consistency procedure needs to be carried out numerically in this case.

The total energy can also be written as
\ben
\label{EAF}
E^{AFM,PM}=\frac{U}{2}\,(1-\frac{\dn^2-\dm^2}{4})-\frac{t^{\rm{eff}}_\uparrow+t^{\rm{eff}}_\downarrow}{2}
\een
In the PM case, the expressions can be simplified to give
\ben
\dn=\frac{2\,\dv-U\,\dn}{\sqrt{\left(\dv-U\dn/2\right)^2+4\,t^2}}
\een
for the occupations and
\ben
\label{EPM}
E^{PM}=\frac{U}{2}\,\left(1-\left(\frac{\dn}{2}\right)^2\right)-\sqrt{\left(\dv-\frac{U}{2}\,\dn\right)^2+4\,t^2}.
\een

\sec{Relation between Hubbard model and Real-space}
\label{app:cnxn}

To show how SOFT and real-space DFT are connected, begin with the one-electron
dimer, H$_2^+$, with the protons separated by $R$.   Use a basis of the exact
atomic 1s orbitals, one on each site.  This is a minimal basis in quantum chemistry.
Then 
\ben
\hat h = -\half \nabla^2 -\frac{1}{r} - \frac{1}{|\br-R{\bf z}|} 
\een
where the bond is along the $z$-axis.  Then the matrix elements of $\hat h$ in
the basis set of atomic orbitals are: 
\ben
v_1= v_2 = \eps_A + j(R),~~ t = s(R)\, \eps_A + k(R)
\een
where $\eps_A$ is the atomic energy (- one Rydberg here) and
\bea
s(R)&=&\langle A | B \rangle = e^{-R} (1 + R + R^2/3)\nonumber\\
j(R)&=&\langle A | \frac{1}{|\br-R{\bf z}|} | A \rangle = -(1/R - e^{-2R}(1 + 1/R))\nonumber\\
k(R)&=&\langle A | \frac{1}{|\br-R{\bf z}|} | B \rangle = - e^{-R}(1+R), 
\eea
yielding the textbook eigenvalues (for the generalized eigenvalue problem):
\ben
\epsilon_\pm = \eps_A + (j\pm k)/(1\pm s).
\een
Of course, the orbitals can always be symmetrically orthogonalized in advance\cite{L50},
in which case 
\bea
v_{\rm{ortho}} &=& \eps_A + (- j + ks)/(s^2 -1),\\
t_{\rm{ortho}} &=& - (s j -k)/(s^2-1).
\eea
Although physics textbooks often set the overlap to zero, this is inconsistent, as the
size of the overlap is comparable to $k(R)$, say.  Setting the on-site potential
to zero (but re-adding its value to the energy) and using $t_{\rm{ortho}}$, makes
the solution Eq. (\ref{e0_N2_U0}) of the text produce the exact electronic energy in this minimal
basis.

But quantum chemistry textbooks note that this calculation is horribly inaccurate,
yielding a bond-length of 2.5 Bohr and a well depth of 2.75 eV.  Inclusion of
a $p_z$ orbital on each site, and allowing the lengthscale of each orbital to
vary, produces almost exact results of 2.00 Bohr and 4.76 eV.  Thus, even in 
this simple case, more than one orbital per site is needed to converge to the
real-space limit.

Next we consider repeating the minimal-basis calculation with one nuclear charge
replaced by value $Z$.  This yields an asymmetric tight-binding problem for which
the orbitals can be orthogonalized and values of $\dv$ and $t$ deduced as a function
of $R$. But note that changing $Z$ will change both $\dv$ and $t$ simultaneously,
unlike our asymmetric SOFT dimer, where only $\dv$ changes.  In real-space DFT, 
the kinetic energy functional remains the same, $T\s\W$ of Eq. (\ref{vonWeisacker}), for all
$R$ and every $Z$.

The situation is even more complicated for H$_2$ and its asymmetric variants.
Clearly $U$ becomes a function of $R$, but there are also several independent off-diagonal
matrix elements that are $R$ dependent.  Again, all change as a function of both
$R$ and $Z$, but none of this occurs in SOFT.   In real-space DFT, $T\s$ is still
the von Weisacker functional, $U_H$ is always the Hartree energy, and the
exact $E\xc[\n]$ is independent of $R$ and $Z$, but always produces the exact energy
when iterated in the KS equations.

%

%\bibliographystyle{apsrmp4-1}
%\bibliographystyle{apsrev4-1}
%\bibliography{Master,hubbard,hubbardspan}

\end{document}